\documentclass[12pt]{article}

\usepackage[USenglish]{babel}

\usepackage[a4paper, left=2.5cm, right=2.5cm, bottom=4.5cm, heightrounded]{geometry}

\usepackage{mathtools,mathrsfs}
\usepackage{amsmath,amsfonts,amssymb}
\usepackage{nccmath}                        %fleqn and other special layouts
\usepackage{upgreek}                        %\uppsi
\usepackage{bbm}                            %\mathbbm - numbers in bb
\usepackage[bbgreekl]{mathbbol}             %greek letters in bb
\usepackage{mathrsfs}                       %\mathscr - cursive letters
\usepackage{mathabx}                        %\widebar and many other math symbols
\usepackage{graphicx}
\usepackage{tikz-cd}
\usepackage{color}
\usepackage{cancel}
\usepackage{slashed}
\usepackage{comment}
\usepackage{cite}
\usepackage[hidelinks]{hyperref}            %links

\numberwithin{equation}{section}

\DeclareSymbolFontAlphabet{\mathbbg}{bbold} %Greek letters in bb
\DeclareSymbolFontAlphabet{\mathbb}{AMSb}   %Latin letters in bb

\newcommand{\Atilde}{{\tilde A}}            %SU(4) index tilde
\newcommand{\Btilde}{{\tilde B}}
\newcommand{\Ctilde}{{\tilde C}}
\newcommand{\Dtilde}{{\tilde D}}
\newcommand{\tmu}{{\tilde\mu}}              %Spacetime tilde
\newcommand{\tnu}{{\tilde\nu}}
\newcommand{\trho}{{\tilde\rho}}

\newcommand{\calK}{\mathcal{K}}             %Graded Vector Space
\newcommand{\calA}{\mathcal{A}}             %Fields
\newcommand{\calE}{\mathcal{E}}             %Equations of Motion
\newcommand{\calN}{\mathcal{N}}             %Noether Identities

\newcommand{\calX}{\mathcal{X}}             %Graded Vector Space
\newcommand{\bbLambda}{\mathbbg{\Lambda}}   %Gauge parameters
\newcommand{\bbH}{\mathbb{H}}               %Fields
\newcommand{\bbE}{\mathbb{E}}               %Equations of Motion
\newcommand{\bbN}{\mathbb{N}}               %Noether Identities
\newcommand{\bbR}{\mathbb{R}}               %NoetherXNoether Identities

\newcommand{\bbB}{\mathbb{B}}               %Operator B
\newcommand{\bbSigma}{\mathbbg{\Sigma}}     %Operator Sigma
\newcommand{\defeq}{{\,\coloneq\,}}         %Definition equal
\newcommand{\del}{\partial}                 %Partial derivative
\newcommand{\delslash}{\slashed{\partial}}  %Partial d. slashed
\newcommand{\Dcov}{\mathcal{D}}             %Covariant derivative
\newcommand{\Tr}{\mathrm{Tr}}               %Trace operator
\newcommand{\1}{\mathbbg{1}}

\def\cA{{\cal A}}
\def\cE{{\cal E}}

\def\cN{{\cal N}}

\def\cK{{\cal K}}

\def\cX{{\cal X}}
\def\cH{{\cal H}}

\def\del{\partial}

\def\Tr{{\rm Tr}}

\def\bpm{\begin{pmatrix}}
\def\epm{\end{pmatrix}}

\allowdisplaybreaks

\begin{document}
\allowdisplaybreaks

\begin{flushright}
    %\today 
    January 3, 2025 \\
    HU-EP-25/02-RTG 
\end{flushright}

\vskip 13mm

\begin{center}
    
    {\large\textbf{The Double Copy of Maximal Supersymmetry in $D=4$}}
    
    \vskip 10mm

    Roberto Bonezzi$^1$, Giuseppe Casale$^2$, Olaf Hohm$^3$

    \vskip 10mm

    {\small\textit{Institut f\"ur Physik, Humboldt-Universit\"at zu Berlin}} \\
    {\small\textit{Zum Großen Windkanal 2, D-12489 Berlin, Germany}}

    \vskip 25mm

    {\bf{Abstract}}

\end{center}

We realize off-shell, local and gauge invariant ${\cal N}=8$ supergravity in $D=4$, to cubic order in fields, as the double copy of ${\cal N}=4$ super Yang-Mills theory (SYM). Employing  the homotopy algebra approach, we show that, thanks to a redundant formulation for the fermionic fields,  the kinematic  algebra ${\cal K}$ of  ${\cal N}=4$ SYM is compatible with an action of the global supersymmetry algebra. The double copy space is then a subspace of ${\cal K}\otimes \widetilde{\cal K}$ that inherits an $L_{\infty}$ algebra on which the two copies of the ${\cal N}=4$ action combine into an action of the ${\cal N}=8$ supersymmetry algebra, with a corresponding enhancement of the $R$-symmetry group to $SU(8)$. 

\vfill

\noindent
{\small
$^1$ bonezzi@physik.hu-berlin.de \\
$^2$ giuseppe.casale@physik.hu-berlin.de \\
$^3$ ohohm@physik.hu-berlin.de
}

\thispagestyle{empty}

\newpage

\clearpage
\pagenumbering{arabic}

\tableofcontents

%%%%%%%%%%%%%%%%%%%%%%%%%%%%%%%%%%
%%%%%%%%%%%%%%%%%%%%%%%%%%%%%%%%
%% BEGIN SECTION 1 (INTRO) %%%%%%%
%%%%%%%%%%%%%%%%%%%%%%%%%%%%%%%%
%%%%%%%%%%%%%%%%%%%%%%%%%%%%%%%%%%
\section{Introduction}
\label{Sec:Intro}

In four spacetime dimensions, which is the physical dimension at least macroscopically, there are two field theories that are distinguished by being maximally symmetric: ${\cal N}=4$ super-Yang-Mills theory (SYM) on the one hand \cite{Brink:1976bc}, and ${\cal N}=8$ supergravity on the other \cite{Cremmer:1979up,deWit:1982bul}.  
${\cal N}=4$ SYM features the maximal number of global supersymmetries in four dimensions, based on the CPT-invariant ${\cal N}=4$ multiplet with maximal spin 1. 
Moreover, ${\cal N}=4$ SYM is a conformally invariant and UV-finite quantum field theory with many further seemingly miraculous properties, including integrability and its role as the  AdS/CFT dual to  type IIB string theory on $AdS_5\times S^5$, see \cite{Brink:2015ust} for  an account of its history. 
Similarly, ${\cal N}=8$ supergravity features the maximal number of local supersymmetries in four dimensions, based on the CPT-invariant ${\cal N}=8$ multiplet with maximal spin 2. 
It is not known whether ${\cal N}=8$ supergravity is UV-finite, but its UV properties are certainly improved, see \cite{Nicolai:2024hqh} for an account  of its history. 
While ${\cal N}=8$ supergravity was at a time considered to be a promising candidate for 
the theory of everything \cite{Hawking:1981sq}, nowadays both theories mostly serve as toy models for quantum field theories in four dimensions that due to their maximal symmetry reveal many remarkable properties. 

One of the most remarkable of these properties is an intimate relation between the two theories: 
${\cal N}=8$ supergravity appears to be, in a suitable sense, the `square' of ${\cal N}=4$ SYM \cite{Bern:2008qj,Bern:2010ue,Carrasco:2011mn,Bern:2011rj,Bern:2019prr,Bern:2022wqg,Adamo:2022dcm}. 
More precisely, the scattering amplitudes of ${\cal N}=4$ SYM may be color-stripped and further manipulated so as to exhibit a property known as color-kinematics duality, in which the color factors (involving structure constants of the gauge group) appear  symmetric with the kinematic factors (functions of momenta and polarization vectors). 
Replacing then the color factors of the ${\cal N}=4$ scattering amplitudes by a second copy of the kinematic factors yields the scattering amplitudes of ${\cal N}=8$ supergravity. This so-called double copy relation has been established at tree-level \cite{Bern:2010yg,Bjerrum-Bohr:2010pnr,Mafra:2011kj}, using its origin in the KLT relations between open and closed string theory \cite{Kawai:1985xq}, and has also been formulated for  pure Yang-Mills theory whose double copy yields the so-called ${\cal N}=0$ supergravity featuring metric, $B$-field and dilaton.  
The double copy is also an important ingredient of the modern revival of higher-loop verifications of UV finiteness of ${\cal N}=8$ supergravity \cite{Bern:2012uf,Bern:2017yxu,Bern:2017ucb}, but complete control over loop-level double copy has not yet been achieved. 

For this and other reasons (such as finding solutions of gravity as the double copy of classical solutions of Yang-Mills theory) one would like to go beyond scattering amplitudes and find an off-shell, local and gauge invariant first-principle derivation of the double copy. We emphasize that here we do not mean `off-shell' in the sense of the supersymmetry transformations closing off shell. 
Rather, we mean that the double copy is performed at the level of the `off-shell' 
theory (as encoded in local and gauge covariant field equations, or ideally an action), as opposed to the `on-shell' scattering amplitudes. 
The resulting supergravities are typically not off-shell in the sense of off-shell closure of the supersymmetry algebra, despite the presence of auxiliary fields.

Recently there has been significant progress in this program, using the framework of homotopy algebras \cite{Borsten:2021hua,Bonezzi:2022yuh,Borsten:2022vtg,Borsten:2022ouu,Bonezzi:2022bse,Bonezzi:2023lkx,Bonezzi:2023pox,Borsten:2023ned,Borsten:2023paw,Bonezzi:2024dlv,Zeitlin:2024cjb} (see also \cite{Anastasiou:2018rdx,Borsten:2020xbt,Borsten:2020zgj,Ferrero:2020vww,Diaz-Jaramillo:2021wtl,Godazgar:2022gfw} for earlier results on off-shell double copy prescriptions). 
One uses that any perturbative semi-classical field theory can be encoded in a homotopy Lie or $L_{\infty}$ algebra \cite{Zwiebach:1992ie,Lada:1992wc,Hohm:2017pnh,Jurco:2018sby} 
whose structure maps or higher brackets encode interaction vertices and other data of the field theory. 
For Yang-Mills theory this $L_{\infty}$ algebra $\cal X_{\rm YM}$ \cite{Zeitlin:2007vv} factorizes into the tensor product of the Lie algebra $\frak{g}$ of the color gauge group and a `kinematic' algebra ${\cal K}$: 
\begin{equation}\label{factorintro}
    \cal X_{\rm YM} = {\cal K}\otimes \frak{g}\;.
\end{equation}  
More precisely, ${\cal K}$ carries the structure of a homotopy commutative associative or $C_{\infty}$ algebra \cite{Zeitlin:2008cc} whose structure maps combine with the Lie algebra structure of $\frak{g}$ to the $L_{\infty}$ brackets of $\cal X_{\rm YM}$. While there is thus a sense in which ${\cal K}$ encodes the pure kinematics of Yang-Mills theory, understanding color-kinematics duality  requires a hidden `Lie-type' structure on ${\cal K}$, which generalizes the $C_{\infty}$  structure and is also a generalization of a Batalin-Vilkovisky (BV) algebra. 
After the early works \cite{Lian:1992mn,Zeitlin:2009tj,Zeitlin:2014xma}, this hidden algebra  was first identified by Reiterer and called BV$_{\infty}^{\Box}$ \cite{Reiterer:2019dys}, referring to new obstructions that arise due to the wave operator $\Box=\partial^{\mu}\partial_{\mu}$ being second-order. 
Using this BV$_{\infty}^{\Box}$ algebra one can then show that on ${\cal K}\otimes \widetilde{\cal K}$, where $\widetilde{\cal K}$ denotes a second copy of the kinematic algebra, there is an $L_{\infty}$ algebra on a certain subspace (the subspace of level-matched states). Starting from pure Yang-Mills theory, this $L_{\infty}$ algebra encodes ${\cal N}=0$ supergravity in a double field theory formulation. 
More precisely, so far this has been established, for manifestly local formulations in generic Lorentzian dimensions, to the order corresponding to quartic couplings \cite{Bonezzi:2022bse}. 
 
Our goal in this paper is to extend these results to ${\cal N}=4$ SYM in order to realize off-shell and gauge invariant ${\cal N}=8$ supergravity as its double copy. Since in ${\cal N}=4$ SYM all fields live in the adjoint representation of the gauge group it is guaranteed  that its $L_{\infty}$ algebra factorizes as in (\ref{factorintro}). 
While it is of course of interest in its own right to work out the details of the kinematic $C_{\infty}$ algebra of ${\cal N}=4$ SYM, in view of double copy the more important question is whether the hidden Lie-type structure and the BV$_{\infty}^{\Box}$ algebra are also realized. We will show here, to the order corresponding to cubic couplings in an action, that this is indeed the case. In order to realize this algebra locally we find it necessary to employ a redundant formulation for the fermionic fields, in which not only the standard Dirac type equations are imposed but also, independently, the second-order Klein-Gordon-type equations that are implied by them. 
Such a fermionic complex is also realized as the Hilbert space of a suitable supersymmetric worldline theory. 
In this formulation one  abandons, for the time being, an action principle, but the advantage is that the so-called $b$ operator that is at the heart of the BV$_{\infty}^{\Box}$ algebra \cite{Ben-Shahar:2021doh,Ben-Shahar:2021zww,Bonezzi:2022yuh,Borsten:2022vtg,Ben-Shahar:2024dju} becomes particularly simple and purely algebraic (not involving any spacetime derivatives). 
 
Given the algebraic structures of ${\cal N}=4$ SYM thus obtained, it is clear from the general results in \cite{Bonezzi:2022yuh,Bonezzi:2022bse} that its double copy, at least to cubic order, yields a manifestly local gravity theory with the dynamics of Einstein-Hilbert type, subject to diffeomorphism gauge invariance (in a double field theory formulation \cite{Siegel:1993th,Hull:2009mi,Hohm:2010jy}). 
However, the most interesting question in the context of  ${\cal N}=4$ SYM is of course whether its supersymmetry gets naturally `doubled' so that its double copy is indeed  ${\cal N}=8$ supergravity. Since the \textit{global} supersymmetry of ${\cal N}=4$ SYM is not automatically encoded in its $L_{\infty}$ algebra, the issue of the double copy of global supersymmetry raises interesting new conceptual questions that we deal with in this paper. 
Specifically, we will show that there is a consistent action of the global supersymmetry algebra, which is of course a strict super-Lie algebra, on the homotopy commutative algebra ${\cal K} = \bigoplus_{i=0}^{\infty} K_i $. 
This action is encoded in $n$-linear maps $ \rho_n(\epsilon): (K_i)^{\otimes\,n} \longrightarrow K_i $  
(of intrinsic degree $1-n$), so that the supersymmetry variations are given by
\begin{equation}
    \delta_\epsilon \cA^a :=
    \rho_1(\epsilon\,|\,\cA^a) + \frac12\,f^a{}_{bc}\,\rho_2(\epsilon\,|\,\cA^b,\cA^c)+\cdots  \,, 
\end{equation} 
where for ease of notation we denote the action of the maps by $x\mapsto \rho_1(\epsilon\,|\,x)$, etc. 
Here $\epsilon$ is the constant supersymmetry parameter, $f^a{}_{bc}$ are the structure constants of the color Lie algebra and $\cA^a\in\cK$ is a field with the color generators removed. 
Note that the maps $\rho_n(\epsilon)$ act in general on rather abstract objects, such as color-stripped equations and Noether identities, not just the fields for which one normally writes down supersymmetry variations. 
The maps $\rho_n(\epsilon)$ satisfy certain 
compatibility conditions with the BV$_{\infty}^{\Box}$ structure maps (for instance, $\rho_1(\epsilon)$ commutes 
with the differential of the $C_{\infty}$ subalgebra). It is intriguing that it is not only possible to display the supersymmetry of ${\cal N}=4$ SYM purely at the level of its kinematic algebra, without any reference to the color gauge group, but that this supersymmetry action is also compatible with the hidden Lie-type structure underlying color-kinematics duality. 

As the second main result of this paper we then show, to cubic order in fields, that the double copy gravity theory on the level-matched subspace of  ${\cal K} \otimes \widetilde{\cal K}$ admits an action of global ${\cal N}=8$ supersymmetry. Specifically, we have now two copies of the global supersymmetry parameter $\epsilon_A$, where $A=1,\ldots, 4$ is the $SU(4)$ $R$-symmetry index, which together combine into the ${\cal N}=8$ supersymmetry parameter 
\begin{equation}
    \epsilon_I \defeq (\epsilon_A, \tilde\epsilon_\Atilde)\,, \quad I=1,\dots,8\;. 
\end{equation}
The supersymmetry action on a generic object of the $L_{\infty}$ algebra is given by $\delta_\epsilon\bbH=\bbSigma_1(\epsilon\,|\,\bbH)+\frac12\,\bbSigma_2(\epsilon\,|\,\bbH,\bbH)+\cdots$, where 
\begin{equation}\label{DCSigma1INTRO}
    \bbSigma_1(\epsilon) \defeq \rho_1(\epsilon)\otimes\tilde{\1} + \1\otimes\tilde{\rho}_1(\tilde\epsilon)\;,
\end{equation}
and $\bbSigma_2$ is given in the main text. 
We have thus realized, at least to cubic order in couplings, ${\cal N}=8$ supersymmetry as the double copy of ${\cal N}=4$. At this stage it should be recalled that the homotopy algebra formulation is perturbative, with field fluctuations around a fixed background, which here we take to be flat Minkowski space. In such a formulation, the more familiar 
\textit{local} supersymmetry of ${\cal N}=8$ supergravity splits into a gauge symmetry with a fermionic gauge parameter and the global ${\cal N}=8$  supersymmetry that leaves the background invariant. 
The former local symmetry is automatically part of the homotopy algebra formulation of double copy, while the latter global symmetry emerges in the novel fashion sketched above. 
We will also show that the manifest $SU(4)\times SU(4)$ $R$-symmetry is enhanced to the $SU(8)$ $R$-symmetry of ${\cal N}=8$ supergravity. 

The rest of this paper is organized as follows. In section~2 we introduce the homotopy algebra formulation of ${\cal N}=4$ SYM, including the supersymmetry action on its kinematic algebra. We then turn in section~3 to its double copy and establish in particular  the double copy of global supersymmetry. 
In section~4 we establish, at the level of the free theory, the relation to ${\cal N}=8$ supergravity in its standard formulation. We conclude in section~5. Our conventions are summarized in appendix \ref{Sec:Conventions}, while appendix \ref{Sec:Fermionic_Complex} discusses the worldline theory underlying the fermionic complex.

%%%%%%%%%%%%%%%%%%%%%%%%%%%%%%%%%%
%%%%%%%%%%%%%%%%%%%%%%%%%%%%%%%%
%% BEGIN SECTION 2 %%%%%%%%%%%%%%%
%%%%%%%%%%%%%%%%%%%%%%%%%%%%%%%%
%%%%%%%%%%%%%%%%%%%%%%%%%%%%%%%%%%
\section{Homotopy Algebra of $\calN=4$ Super Yang-Mills}
\label{Sec:Hom_Alg_Of_SYM}

In this section we begin by providing the action for $\calN=4$ super Yang-Mills (SYM) theory, together with its gauge and supersymmetry transformations. Thereafter, the theory will be reformulated and analyzed within the framework of homotopy algebras, both in its standard and color-stripped versions. This will provide all the ingredients necessary to apply the off-shell double copy prescription developed in \cite{Bonezzi:2022yuh,Bonezzi:2022bse}. The last part of this section will be devoted to the reformulation of global supersymmetry as an action on the homotopy algebra of the theory.

%%%%%%%%%%%%%%%%%%%%%%%%%%%%%%%%%
%%%%%%%%%%%%%%%%%%%%%%%%%%%%%%%%
%%%%%%%%%%%%%%%%%%%%%%%%%%%%%%%%%%
\subsection{$\calN=4$ super Yang-Mills: Action and Symmetries}
\label{Subsec:N=4SYM_Action_Sym}

We begin by displaying the full action for $\calN=4$ super Yang-Mills theory \cite{Brink:1976bc}: 
\begin{equation}\begin{split}
    S_{\mathrm{SYM}} = \int d^4x\, \Tr\,\Big\{ &-\frac14 F_{\mu\nu}F^{\mu\nu} + \frac12\Dcov_\mu\phi_{AB}\Dcov^\mu\phi^{AB}  + i\,\widebar\chi^A\widebar\sigma^\mu\Dcov_\mu\chi_A+ \\
    &+ \widebar\chi^A[\phi_{AB},\widebar\chi^B] + \chi_A[\phi^{AB},\chi_B] - \frac14[\phi_{AB},\phi_{CD}][\phi^{AB},\phi^{CD}] \Big\}\;,
    \label{Action:SYM_full}
\end{split}\end{equation}
where $\mu=0,\dots,3$ is a spacetime Lorentz index and $A=1,\dots,4$ is a fundamental index of $ \mathrm{SU}(4)$. All fields are valued in the color Lie algebra,  with the covariant derivative defined as $\Dcov_\mu = \partial_\mu + [A_\mu,-]$. The scalars $\phi_{AB}$, $\phi^{AB}$ are subject to the reality conditions
\begin{equation}
    \phi^{AB} = \frac12\,\epsilon^{ABCD}\phi_{CD} = (\phi_{AB})^*\;, 
    \label{Eq:Reality_Cond_Scalars}
\end{equation}
and transform in the representation $\mathbf{6}$ (or $\mathbf{\bar6}$, respectively) of $\mathrm{SU}(4)$. 
The left- and right-handed Weyl spinors\footnote{Later on we will omit the spinor indices $\alpha=1,2$ and $\dot\alpha=\dot1,\dot2$.} (gaugini) $\chi_{A\alpha}$ and $\widebar\chi^A_{\dot\alpha}$ satisfy the reality condition
\begin{equation}
    \widebar\chi^A_{\dot\alpha} = (\chi_{A\alpha})^\dagger\;, 
    \label{Eq:Reality_Cond_Spinors}
\end{equation}
and transform in the $\mathbf{4}$ and in the $\mathbf{\bar4}$ representations of $\mathrm{SU}(4)$.

The action \eqref{Action:SYM_full} is invariant under both the gauge transformations
\begin{subequations}\begin{align}
    \delta_\Lambda A_\mu &= \Dcov_\mu\Lambda\;,  \\[1mm]
    \delta_\Lambda \phi_{AB} &= [\phi_{AB},\Lambda]\;,  \\[1mm]
    \delta_\Lambda \phi^{AB} &= [\phi^{AB},\Lambda] \;, \\[1mm]
    \delta_\Lambda \chi_A &= [\chi_A,\Lambda]\;,  \\[1mm]
    \delta_\Lambda \widebar\chi^A &= [\widebar\chi^A,\Lambda]\;, 
\end{align}\end{subequations}
and global supersymmetry transformations
\begin{subequations}\begin{align}
    \delta_\epsilon A_\mu &= \epsilon_A\sigma_\mu\widebar\chi^A \;, \\[1mm]
    \delta_\epsilon \phi_{AB} &= 2i\,\epsilon_{[A}\chi_{B]}\;,  \\[1mm]
    \delta_\epsilon \phi^{AB} &= i\,\epsilon^{ABCD}\epsilon_{C}\chi_{D}\;,  \\[1mm]
    \delta_\epsilon \chi_A &= \epsilon_A\sigma^{\mu\nu} F_{\mu\nu} - 2i\,[\phi_{AC},\phi^{CB}]\epsilon_B\;,  \\[1mm]
    \delta_\epsilon \widebar\chi^A &= 2\,\epsilon_B\sigma^\mu\Dcov_\mu\phi^{AB} \;,
\end{align}\end{subequations}
where we omitted the hermitian conjugate terms transforming with $\widebar\epsilon^A$.
Moreover, the action yields  the following equations of motion:
\begin{subequations}\begin{align}
    \Dcov^\nu F_{\nu\mu} + [\phi^{AB},\Dcov_\mu\phi_{AB}] - i\, [\widebar\chi^A,\widebar\sigma_\mu\chi_A] &= 0\;,  \\[1mm]
    -\Dcov^\mu\Dcov_\mu\phi^{AB} - [\widebar\chi^A,\widebar\chi^B] - \frac12\epsilon^{ABCD}[\chi_C,\chi_D] + [[\phi^{AB},\phi^{CD}],\phi_{CD}] &= 0\;,  \\[1mm]
    i\,\widebar\sigma^\mu\Dcov_\mu\chi_A +2\,[\phi_{AB},\widebar\chi^B] &= 0\;,  \\[1mm]
    i\,\sigma^\mu\Dcov_\mu\widebar\chi^A + 2\,[\phi^{AB},\chi_B]&= 0 \,.
    \label{Eq:SYM_eoms}
\end{align}\end{subequations}

For reasons that will be clear in the following, it is convenient to expand the action explicitly in powers of the fields
\begin{equation}\begin{split}
    S_{\mathrm{SYM}} = \int d^4x\, \Tr\,&\Big\{ \frac12 A_\mu\Box A^\mu + \frac12(\del\!\cdot\!A)^2 - \frac12\phi^{AB}\Box\,\phi_{AB} + i\,\widebar\chi^A\widebar\sigma^\mu\del_\mu\chi_A + \\
    &- \del_\mu A_\nu[A^\mu,A^\nu] + \del_\mu\phi_{AB}[A^\mu,\phi^{AB}] + i\,\widebar\chi^A\widebar\sigma^\mu[A_\mu,\chi_A] + \\
    &+ \widebar\chi^A[\phi_{AB},\widebar\chi^B] + \chi_A[\phi^{AB},\chi_B] + \\
    &- \frac14[A_\mu,A_\nu][A^\mu,A^\nu] + \frac12[A_\mu,\phi_{AB}][A^\mu,\phi^{AB}] + \\
    &- \frac14[\phi_{AB},\phi_{CD}][\phi^{AB},\phi^{CD}] \Big\}\,,
    \label{Action:SYM_full_explicit}
\end{split}\end{equation}
where in the first line one reads off  the free theory, followed by the cubic and quartic interaction terms. The explicit perturbative expansion hides the gauge covariance, but will allow us to read off the homotopy algebraic structure of the theory, to which we turn next.

%%%%%%%%%%%%%%%%%%%%%%%%%%%%%%%%%
%%%%%%%%%%%%%%%%%%%%%%%%%%%%%%%%
%%%%%%%%%%%%%%%%%%%%%%%%%%%%%%%%%%
\subsection{$L_\infty$ Formulation}
\label{Subsec:Linf_Form}

In general, every perturbative field theory admits a description in terms of \emph{homotopy Lie algebras} -- also known as $L_\infty$-algebras, 
see e.g.~\cite{Hohm:2017pnh}. 
This mathematical 
structure consists of the data $(\calX,B_n)$ of a graded vector space 
\begin{equation}
    \calX \defeq \bigoplus_{i\in\mathbb{Z}} X_i\;,
\end{equation}
together with graded symmetric multilinear maps 
\begin{equation}
    B_n: \, \calX^{\otimes n} \longrightarrow \calX, \quad n\geq 1\;,
\end{equation}
of intrinsic degree $|B_n|=+1$.
These encode the perturbative structure 
of a field theory and are subject to 
generalized Jacobi identities, 
including that 
$B_1$ is a nilpotent operator. 
Given the nilpotent differential $B_1$, one can organize the graded vector space $\cX$ into a chain complex
\begin{equation}
   \begin{tikzcd}[row sep=0pt, column sep={2cm,between origins}]
        \cdots \arrow[r,"B_1"] & X_{-1} \arrow[r,"B_1"] & X_{0} \arrow[r,"B_1"]  &X_{1}\arrow[r,"B_1"] & X_{2}\arrow[r,"B_1"] & \cdots \\
        & \Uplambda & \calA & \calE & \calN &
   \end{tikzcd}
\end{equation}
with the elements $\Uplambda$, $\calA$, $\calE$, $\calN$ in increasing degree being respectively the sets of gauge parameters, fields, equations of motion and Noether identities. The brackets $B_n(\cA,\ldots,\cA)$ between fields describe the interactions. Specifically, upon introducing  an inner product $\langle\cdot,\cdot\rangle$ the perturbative action takes the generalized Maurer-Cartan form
\begin{equation}
    S = \sum_{n=1}^{\infty}\frac1{(n+1)!}\langle\calA,B_n(\underbrace{\calA,\dots,\calA}_{\text{$n$ times}})\rangle \,.
    \label{Action:L_infty}
\end{equation}
Given the inner product, each $n$-linear map $B_n$ provides all the information about the $(n+1)$th order of interaction in the action. In particular, the brackets $B_n(\cA,\ldots,\cA)$ themselves correspond to the perturbative expansion of the equations of motion, in that
\begin{equation}\label{MC equation}
    \frac{\delta S}{\delta\calA} = \sum_{n=1}^{\infty}\frac1{n!} B_n(\underbrace{\calA,\dots,\calA}_{\text{$n$ times}}) \;.
\end{equation}
Similarly, gauge transformations are related to the brackets $B_n(\Uplambda,\cA,\ldots,\cA)$ via the expansion of $\delta_\Uplambda\cA$ in powers of fields as $\delta_{\Uplambda}\cA=B_1(\Uplambda)+B_2(\Uplambda,\cA)+\cdots$.
For a more in-depth description of the $L_\infty$ formulation of gauge theories, we refer to \cite{Hohm:2017pnh,Jurco:2018sby}.

We will now specialize to the case of ${\calN=4}$ super Yang-Mills theory. In view of performing a local double copy, we will consider a modified chain complex to describe the free theory. To begin with, following \cite{Bonezzi:2022yuh} we add an auxiliary scalar field $\varphi$ and use the modified quadratic action
\begin{equation}
    S_{\mathrm{SYM}}^{\textit{free}} = \int d^4x\,\Tr \left\{ \frac12A_\mu\Box A^\mu - \frac12\varphi^2 + \del\!\cdot\!A\,\varphi - \frac12\phi^{AB}\Box\,\phi_{AB} + i\,\widebar\chi^A\widebar\sigma^\mu\del_\mu\chi_A \right\}.
    \label{Action:SYM_free+aux}
\end{equation}
The extra field $\varphi$ can be integrated out algebraically by setting $\varphi = \del\cdot A$, which reproduces the standard free action. To ensure gauge invariance, we declare $\varphi$ to transform as its on-shell value: $\delta_\Lambda\varphi\equiv\del_\mu(\delta_\Lambda A^\mu) = \Box\Lambda + \del_\mu[A^\mu,\Lambda]$.

The chain complex associated with the free theory is given by the direct sum of the bosonic and fermionic subspaces: $\cX_{\rm SYM}=\cX^{\rm B}\oplus\cX^{\rm F}$. The bosonic complex consists of the spaces of gauge parameters, bosonic fields, their equations of motion and Noether identities as follows
\begin{equation}
\label{BoseChain}
   \begin{tikzcd}[row sep=0pt, column sep={2.5cm,between origins}]
        X^{\rm B}_{-1} \arrow[r,"B_1"] & X^{\rm B}_{0} \arrow[r,"B_1"]  &X^{\rm B}_{1}\arrow[r,"B_1"] &X^{\rm B}_{2} \\
        \Lambda & A_\mu, \phi_{AB}& E &  \\
        & \varphi & E_\mu, E_{AB} & N
   \end{tikzcd}\;,
\end{equation}
where $E$, $E_\mu$ and $E_{AB}$ are the equations of motion of $\varphi$, $A_\mu$ and $\phi_{AB}$, respectively. The fermionic chain complex, on the other hand, consists only of the space of fermionic fields and their equations:
\begin{equation}
\label{FermiChain}
   \begin{tikzcd}[row sep=0pt, column sep={2.5cm,between origins}]
        0 \arrow[r] & X^{\rm F}_{0} \arrow[r,"B_1"]  &X^{\rm F}_{1}\arrow[r] &0 \\
        &\chi_A& \cE_A &  
   \end{tikzcd}\;.
\end{equation}
The equation for the left-handed Weyl spinor $\chi_{A\alpha}$ is the right-handed Weyl spinor $\cE_A^{\dot\alpha}$ and above we omitted the complex conjugate $\bar\chi^A$ together with its equation $\bar\cE^A$.
The differential $B_1$ acts separately on $\cX^{\rm B}$ and $\cX^{\rm F}$. The linearized gauge transformations are given by $\delta_\Lambda\cA=B_1(\Lambda)$, with
\begin{equation}
    B_1(\Lambda) = \bpm \del_\mu\Lambda \,,\; 0 \\[1mm] \Box\Lambda \epm \,\in X_0^{\rm B} \;.    
\end{equation}
One can see that $B_1(\Lambda)$ only takes values in the space $X_0^{\rm B}$ of bosonic fields, as only $A_\mu$ and $\varphi$ transform to lowest order. Acting with the differential on fields yields the linearized equations of motion $B_1(\cA)=0$, where
\begin{equation}
    B_1 \bpm A_\mu\,,\;\phi_{AB} \\[1mm] \varphi \epm =
    \bpm\del\cdot A - \varphi \\[1mm] \Box A_\mu - \del_\mu\varphi \,,\; -\Box\phi_{AB}\epm\,\in\,X_1^{\rm B}\;, \quad
    B_1(\chi_A) = i\bar\sigma^\mu\del_\mu\chi_A\,\in\,X_1^{\rm F}\;.     
\end{equation}
Finally, the Noether identity is encoded in $B_1(\cE)$, which acts non-trivially only on the bosonic equations of $A_\mu$ and $\varphi$:
\begin{equation}
    B_1 \bpm E \\[1mm] E_\mu \,,\; E_{AB}\epm = \Box E-\del^\mu E_\mu\,\in\,X_2^{\rm B}\;, \quad B_1(\cE_A) = 0\;.    
\end{equation}
The nilpotence of the differential expresses gauge invariance of the free field equations as $B_1^2(\Lambda)\equiv0$, as well as the linearized Noether identity $B_1^2(\cA)\equiv0$.

At this stage, the fermionic complex $\cX^{\rm F}$ is not suitable to apply the double copy procedure of \cite{Bonezzi:2022yuh,Bonezzi:2022bse}. The reason is that the method developed in \cite{Bonezzi:2022yuh,Bonezzi:2022bse} requires a second differential of degree $-1$, named $b$, defined by the properties
\begin{equation}\label{bdefining}
    b^2=0\;, \qquad b\,B_1+B_1\,b=\Box \;,   
\end{equation}
which, in some sense, provides an invariant definition of the wave operator $\Box=\del^\mu\del_\mu$. In the bosonic complex \eqref{BoseChain} such $b$ operator is given by a simple degree shift, acting as follows
\begin{subequations}\begin{align}
    b\,(N) &= \bpm N \\[1mm] 0 \,,\; 0 \epm \,\in\, X_1^{\rm B} \;, \\[3mm]
    b\bpm E \\ E_\mu \,,\; E_{AB} \epm &= \bpm E_\mu \,,\; -E_{AB} \\[1mm] 0 \epm \,\in\, X_0^{\rm B}\;, \\[3mm]
    b\bpm A_\mu \,,\; \phi_{AB} \\[1mm] \varphi \epm &= \varphi \,\in\, X_{-1}^{\rm B}\;,    
\end{align}\end{subequations}
which can be visualized by the diagram
\begin{equation}
\label{Bosebdiagram}
   \begin{tikzcd}[row sep=0pt, column sep={2.5cm,between origins}]
        X^{\rm B}_{-1} \arrow[r,"B_1"] & X^{\rm B}_{0} \arrow[r,"B_1"]  &X^{\rm B}_{1}\arrow[r,"B_1"] &X^{\rm B}_{2} \\
        \Lambda & A_\mu, \phi_{AB}& E &  \\
        & \arrow[ul,"b"]\varphi &\arrow[ul,"b"] E_\mu, E_{AB} &\arrow[ul,"b"] N
   \end{tikzcd}\;.
\end{equation}
In particular, the introduction of the auxiliary scalar $\varphi$ is instrumental for the $b$ operator to be local and, furthermore, not to contain spacetime derivatives.

In the fermionic complex \eqref{FermiChain}, an operator satisfying \eqref{bdefining} can be defined by 
\begin{equation}\label{Naiveb}
b(\cE_A)=-i\sigma^\mu\del_\mu\cE_A\,\in\,X_0^{\rm F}\;,\quad b(\chi_A)=0\;.    
\end{equation}
However, the fields of the double copy theory are subject to constraints involving the $b$ operator, which we will discuss in the next section. Choosing \eqref{Naiveb} would thus lead to fields in the double copy subject to differential constraints, which we prefer to avoid.
A possible way to circumvent such problem is to redefine the differential $B_1$ on the fermionic sector, by adding a second component to the fermion equation: 
\begin{equation}
    B_1(\chi_A) = \bpm i\widebar\sigma^\mu\del_\mu\chi_A \\[1mm] \Box\chi_A \epm \,\in\,X_1^{\rm F}\;,
\end{equation}
so that the space $X_1^{\rm F}$ consists of the doublet of spinors $(\cE_A^{\dot\alpha}, E_{A\alpha})$. One can now define the $b$ operator as the degree shift
\begin{equation}\label{Fermib}
    b \bpm \cE_A \\[1mm] E_A \epm = E_A\,\in\,X_0^{\rm F}\;, \quad b(\chi_A)=0\;.    
\end{equation}
The second equation $E_A$ is completely redundant, in that $\Box\chi_A=-i\sigma^\nu\del_\nu(i\bar\sigma^\mu\del_\mu\chi_A)$. Adding it to the complex does not change the dynamics in any way, but it introduces an infinite series of artificial Noether identities, due to the fact that the two equations are not independent. The resulting fermionic complex becomes semi-infinite: $\cX^{\rm F}=\bigoplus_{i=0}^{\infty}X_i^{\rm F}$, with the differential $B_1$ extended as in the following diagram
\begin{equation}
\label{CC:SYM_Linf_ferm}
   \begin{tikzcd}[row sep=10pt, column sep={2.5cm,between origins}]
        0 \arrow[r] & X_{0}^{\rm F} \arrow[r,"B_1"]  &X_{1}^{\rm F}\arrow[r,"B_1"] &X_{2}^{\rm F}\arrow[r,"B_1"] & \cdots \\
        & \chi_A \arrow[r,"i\widebar\sigma^\mu\del_\mu"]\arrow[dr,"\Box"] & \calE_A \arrow[r,"i\sigma^\nu\del_\nu"] & \calN_A \arrow[r,"i\widebar\sigma^\rho\del_\rho"]\arrow[dr,"\Box"] & \cdots \\
        & & E_A \arrow[ru,"1"]\arrow[r,"-i\widebar\sigma^\mu\del_\mu"] & N_A \arrow[r,"-i\sigma^\nu\del_\nu"] & \cdots
   \end{tikzcd}
\end{equation}
Despite introducing infinitely many trivial Noether identities, this formulation allows us to define the $b$ operator as a degree shift on the whole complex. On $\cX^{\rm F}$ this is given by
\begin{equation}
\label{Fermibdiagram}
   \begin{tikzcd}[row sep=10pt, column sep={2.5cm,between origins}]
        0 \arrow[r] & X_{0}^{\rm F} \arrow[r,"B_1"]  &X_{1}^{\rm F}\arrow[r,"B_1"] &X_{2}^{\rm F}\arrow[r,"B_1"] & \cdots \\
        & \chi_A & \calE_A  & \calN_A & \cdots \\
        & &\arrow[ul,"b"]E_A & \arrow[ul,"b"]N_A  & \arrow[ul,"b"]\cdots
   \end{tikzcd}\;,
\end{equation}
and will result in a local double copy in terms of unconstrained fields. This reformulation of the fermionic $L_\infty$ complex arises  from the BRST quantization of a spinning worldline theory, which we discuss in appendix \ref{Sec:Fermionic_Complex}.

%%%%%%%%%%%%%%%%%%%%%%%%%%%%%%%%%
%%%%%%%%%%%%%%%%%%%%%%%%%%%%%%%%
%%%%%%%%%%%%%%%%%%%%%%%%%%%%%%%%%%
\subsection{Kinematic Algebra}
\label{Subsec:Kin_Alg}

Bearing in mind that the aim is to perform the double copy of $\calN=4$ super Yang-Mills theory, we need to `color-strip' the $L_\infty$ algebra $(\cX_{\rm SYM}, B_n)$ and then take  the tensor product of the resulting structure with a second copy of itself. In a theory where all fields and gauge parameters take value in a color Lie algebra $\mathfrak{g}$, the $L_\infty$ vector space $\cX$ can be written as the tensor product $\cX=\cK\otimes\mathfrak{g}$. The graded vector space $\cK$, whose elements have no color degrees of freedom, then carries a kinematic $C_\infty$ algebra, with products $\{m_n\}_{n\geq1}$, which is a homotopy generalization of associative commutative algebras.
The products $m_n$ are $n$-linear maps
\begin{equation}
    m_n: \, \calK^{\otimes n} \longrightarrow \calK \;, \quad|m_n| = 2-n\;,
    \label{Def:C_inf_nmaps}
\end{equation}
obeying generalized associativity conditions \cite{Borsten:2021hua,Bonezzi:2022yuh}.

The $C_\infty$ algebra $(\cK_{\rm SYM},m_n)$ of $\cN=4$ SYM theory has non-vanishing products $m_1$, $m_2$ and $m_3$. $m_1$ is a nilpotent differential that obeys the Leibniz rule with respect to the product $m_2$. The two-product, in turn, is associative up to homotopy, given by $m_3$. The $C_\infty$ products can be read off from the $L_\infty$ brackets $B_n$ by means of the relations \cite{Bonezzi:2023ciu}
\begin{fleqn}
\begin{subequations}\begin{align}
    B_1(x) &= m_1(u^a)\otimes t_a \;, \\[1mm]
    B_2(x_1,x_2) &= (-1)^{x_1}f^a{}_{bc}\,m_2(u_1^b,u_2^c)\otimes t_a \;, \\[1mm]
    B_3(x_1,x_2,x_3) &= f^a{}_{be}f^e{}_{cd}\,\Big[(-1)^{x_2}m_3(u_{1}^b,u_2^c,u_{3}^d)\! +\! (-1)^{x_1(x_2+1)}m_3(u_{2}^b,u_1^c,u_{3}^d)\Big]\! \otimes t_a \,,
\end{align}\label{Eq:Linf_Cinf_relations}\end{subequations}
\end{fleqn}
where $t_a$ and $f^a{}_{bc}$ are generators and structure constants of $\mathfrak{g}$.
Here $x_i\in\calX_{\rm SYM}$ are arbitrary elements of the $L_\infty$ vector space. They have been expanded as $x_i=u_i^a\otimes t_a$, with $u_i^a\in\calK_{\rm SYM}$. The degrees in $\cX_{\rm SYM}$ and $\cK_{\rm SYM}$ are related via $|x|=|u|-1$ for $x=u\otimes t$. The $C_\infty$ chain complex is also given by the direct sum of the bosonic and fermionic sectors: $\cK_{\rm SYM}=\cK^{\rm B}\oplus\cK^{\rm F}$. Together, they form the semi-infinite complex
\begin{equation}
\label{CC:SYM_Cinf_full}
   \begin{tikzcd}[row sep=0pt, column sep={2.5cm,between origins}]
        K_{0} \arrow[r,"m_1"] & K_{1} \arrow[r,"m_1"]  &K_{2}\arrow[r,"m_1"] &K_{3}\arrow[r,"m_1"] & \cdots \\
        \Lambda & A_\mu, \phi_{AB}, \chi_A & E, \calE_A & \calN_A & \cdots \\
        & \varphi & E_\mu, E_{AB}, E_A & N, N_A & \cdots
   \end{tikzcd}
\end{equation}
where we use the same symbols for the color-stripped elements.
Notice that, since $B_1$ acts trivially on the color Lie algebra, its explicit expression coincides with $m_1$. 

It turns out that the $C_\infty$ algebra obtained by color-stripping is only the superficial layer of a much larger kinematic algebra \cite{Reiterer:2019dys,Bonezzi:2022bse,Borsten:2022vtg,Bonezzi:2023pox,Borsten:2023ned,Bonezzi:2023lkx}, which governs the double copy. 
This includes the $b$ operator introduced in the previous subsection, which obeys the defining properties
\begin{equation}
  b^2=0\;, \qquad  b\,m_1+m_1\,b = \Box\;,  \qquad |b|=-1\;.
    \label{Prop:BV_inf_box}
\end{equation}
Just as $B_1$, the $b$ operator does not act on the color Lie algebra, so that its action on the kinematic complex $\cK_{\rm SYM}$ is the same degree shift we introduced previously.
Instead of being a derivation of the product $m_2$, the failure of $b$ to obey the Leibniz rule with respect to $m_2$ defines a kinematic bracket $b_2$ via
\begin{equation}
b_2(u,v)\defeq bm_2(u,v)-m_2(bu,v)-(-1)^um_2(u,bv)\;.    
\end{equation}
The bracket $b_2$, which is defined similarly to the antibracket of the BV formalism, is the starting point for this homotopy kinematic algebra, termed BV$_\infty^\Box$ in \cite{Reiterer:2019dys}, which is at the core of the off-shell double copy program.

%%%%%%%%%%%%%%%%%%%%%%%%%%%%%%%%%
%%%%%%%%%%%%%%%%%%%%%%%%%%%%%%%%
%%%%%%%%%%%%%%%%%%%%%%%%%%%%%%%%%%
\subsection{Global $\calN=4$ Supersymmetry}
\label{Sec:SYM_Global_Susy}

Apart from some technical details, such as the modified fermionic chain complex, the homotopy algebra formulation of $\cN=4$ super Yang-Mills theory is essentially the same as its bosonic counterpart. The novel feature of $\cN=4$ is global supersymmetry and, in particular, understanding its fate under the double copy procedure.
In this subsection we reformulate global $\calN=4$ supersymmetry in an algebraic language, in compliance with what has been done in the previous subsections. Again, we are going to focus on both standard and color-stripped SYM, the latter being essential for the construction of the double copy. This process will shed light on how supersymmetry acts on the kinematic algebra of the gauge theory.

To start, let us recall that the free action \eqref{Action:SYM_free+aux} is invariant under the following linearized supersymmetry transformations:
\begin{subequations}\label{Eq:SYM_SUSY_transf_linear}\begin{align}
    \delta_\epsilon A_\mu &= \epsilon_A\sigma_\mu\widebar\chi^A\;, \\[1mm]
    \delta_\epsilon\varphi &= \epsilon_A\sigma_\mu\del^\mu\widebar\chi^A \;, \\[1mm]
    \delta_\epsilon \phi_{AB} &= 2i\,\epsilon_{[A}\chi_{B]}\;, \\[1mm]
    %\delta_\epsilon \phi^{AB} &= i\,\epsilon^{ABCD}\epsilon_{C}\chi_{D}\;, \\[1mm]
    \delta_\epsilon \chi_A &= 2\,\epsilon_A\sigma^{\mu\nu}\del_\mu A_\nu \;, \\[1mm]
    \delta_\epsilon \widebar\chi^A &= 2\,\epsilon_B\sigma^\mu\del_\mu\phi^{AB} \;.
\end{align}\end{subequations}
The transformation rule for the auxiliary field $\varphi$ has been chosen to coincide with the one of its on-shell value: $\delta_\epsilon\varphi=\del^\mu\delta_\epsilon A_\mu$, which ensures that the action remains invariant.
As a consequence, the free equations of motion
%\begin{gather*}
 %   E = \del\!\cdot\!A - \varphi \,, \quad
  %  \calE_A = i\widebar\sigma^\mu\del_\mu\chi_A \,, \\
   % E_\mu = \Box A_\mu - \del_\mu\varphi \,, \quad
   % E_i = -\Box\phi_i \,, \quad
   % E_A = \Box\chi_A
%\end{gather*}
are rotated into one another under linearized supersymmetry. Explicitly, we have
\begin{subequations}\label{deltaequations}\begin{align}
    \delta_\epsilon E &= 0\;, \\[1mm]
    \delta_\epsilon E_\mu &= \epsilon_A(\sigma_\mu\widebar E^A +i \del_\mu\widebar{\calE}^A)\;, \\[1mm]
    \delta_\epsilon E_{AB} &= -2i\,\epsilon_{[A}E_{B]} \;, \\[1mm]
    %\delta_\epsilon E^{AB} &= -i\,\epsilon^{ABCD}\epsilon_C E_D \\[1mm]
    \delta_\epsilon \calE_A &= (E_\mu - \del_\mu E)\widebar\sigma^\mu\epsilon_A\;, \\[1mm]
    \delta_\epsilon E_A &= - 2\,\del_\mu E_\nu\,\sigma^{\mu\nu}\epsilon_A+i\,(\Box E-\del^\mu E_\mu)\,\epsilon_A \;, \\[1mm]
    \delta_\epsilon \widebar{\calE}^A &= 2i\, \epsilon_BE^{AB} \;, \\[1mm]
    \delta_\epsilon \widebar{E}^A &= 2\,\widebar\sigma^\mu\epsilon_B \del_\mu E^{AB}\;,
\end{align}\end{subequations}
where we recall that $E_\mu$ and $E$ are the equations for $A_\mu$ and $\varphi$ and $E_{AB}$ is the equation for the matter scalars $\phi_{AB}$. For fermions, $\cE_A$ and $\bar\cE^A$ are the equations of $\chi_A$ and $\bar\chi^A$, while $E_A$ and $\bar E^A$ are their dependent second order equations introduced previously. 
As it is expected, all the equations of motion of the bosonic fields are rotated into equations of motion for the fermionic fields and vice versa. Covariance of the field equations under supersymmetry is a weaker requirement than invariance of the action. We will exploit this later on, when discussing the action of supersymmetry on the kinematic algebra.

At the level of the $L_\infty$ algebra we can define some supersymmetry $n$-linear maps
\begin{equation}
    \Sigma_{n}(\epsilon):\;\cX_{\rm SYM}^{\otimes\,n} \to \cX_{\rm SYM}\;, \quad |\Sigma_n(\epsilon)|=0\;,
\end{equation}
$\epsilon_A$ being the global supersymmetry parameter. Their action on fields is defined by the perturbative expansion of the supersymmetry variations
\begin{equation}
    \delta_\epsilon\,\cA=: \Sigma_1(\epsilon\,|\,\cA) + \frac12\Sigma_2(\epsilon\,|\,\cA,\cA)+\cdots \,, \quad \cA\,\in X_0\;.
\end{equation}
Similarly, covariance of the field equations is expressed via
\begin{equation}
\delta_\epsilon\cE=: \Sigma_1(\epsilon\,|\,\cE) + \Sigma_2(\epsilon\,|\,\cE,\cA)+\cdots \,, \quad \cA\,\in X_0\;,\;\cE\,\in X_1\;.    
\end{equation}
%On the chain complex, this can be visualized as follows:
%\begin{equation}
%    \begin{tikzcd}[row sep=1cm, column sep={2cm,between origins}]
%        &\Uplambda \arrow[r,"B_1"]\arrow[d,"\Sigma_1"] &\calA \arrow[r,"B_1"]\arrow[d,"\Sigma_1"] &\calE \arrow[r,"B_1"]\arrow[d,"\Sigma_1"] &\calN \arrow[r,"B_1"]\arrow[d,"\Sigma_1"] & \cdots\arrow[d,"\Sigma_1"] \\
        %%%
%        &\delta_\epsilon\Uplambda \arrow[r,"B_1"] &\delta_\epsilon\calA \arrow[r,"B_1"] &\delta_\epsilon\calE \arrow[r,"B_1"] 
 %       &\delta_\epsilon\calN \arrow[r,"B_1"] & \cdots
 %  \end{tikzcd}\hspace{1.4cm}
%\end{equation}
More generally, supersymmetry of the theory at linear order is expressed by demanding that $\Sigma_1(\epsilon)$ commutes with the $L_\infty$ differential:  
\begin{equation}
    B_1 \circ \Sigma_1(\epsilon) = \Sigma_1(\epsilon) \circ B_1 \;,
    \label{Prop:SYM_GlobalSUSY1}
\end{equation}
or, in short, ${[B_1,\Sigma_1(\epsilon)]=0}$.
Using \eqref{Eq:SYM_SUSY_transf_linear} for $\Sigma_1(\epsilon\,|\,\cA)$ and \eqref{deltaequations} for $\Sigma_1(\epsilon\,|\,\cE)$, such requirement defines the action of the map $\Sigma_1(\epsilon)$ recursively on the whole complex. 

At quadratic order in fields, imposing covariance under supersymmetry of the equations of motion, written in the Maurer-Cartan form \eqref{MC equation}, amounts to requiring
\begin{equation}\begin{split}
    \delta_\epsilon\big(B_1(\cA) + \tfrac12 B_2(\cA,\cA)\big) &= B_1\big(\Sigma_1(\epsilon\,|\,\cA) + \tfrac12 \Sigma_2(\epsilon\,|\,\cA,\cA)\big) + B_2(\Sigma_1(\epsilon\,|\,\cA),\cA) \\
    &\stackrel{!}{=}\Sigma_1\big(\epsilon\,|\,B_1(\cA)+\tfrac12\,B_2(\cA,\cA)\big)+\Sigma_2\big(\epsilon\,|\,B_1(\cA),\cA\big)\;.
\end{split}\end{equation}
Given the linear relation $[B_1,\Sigma_1(\epsilon)]=0$, this holds as long as
\begin{equation}
     B_1\Sigma_2(\epsilon\,|\,\cA,\cA) - 2\,\Sigma_2(\epsilon\,|\,B_1\cA,\cA) = \Sigma_1\big(\epsilon\,|\,B_2(\cA,\cA)\big) - 2\,B_2\big(\Sigma_1(\epsilon\,|\,\cA),\cA\big)\;.
\end{equation}
In general, this suggests that a consistent action of supersymmetry on the $L_\infty$ algebra requires $[B_1,\Sigma_2(\epsilon)]=[\Sigma_1(\epsilon),B_2]$ or, explicitly,
\begin{equation}\label{Prop:SYM_GlobalSUSY2}\begin{split}
    &B_1\Sigma_2(\epsilon\,|\,x,y)-\Sigma_2(\epsilon\,|\,B_1x,y)-(-1)^x\Sigma_2(\epsilon\,|\,x,B_1y) \\
    &=\Sigma_1\big(\epsilon\,|\,B_2(x,y)\big)-B_2\big(\Sigma_1(\epsilon\,|\,x),y\big)-B_2\big(x,\Sigma_1(\epsilon\,|\,y)\big)\;,
\end{split}\end{equation}
for $x$ and $y$ arbitrary elements of $\cX_{\rm SYM}$. One can continue to the next order, thereby obtaining a compatibility relation between $\Sigma_1(\epsilon)$, $B_3$, $\Sigma_2(\epsilon)$ and $B_2$, but we will not pursue this here.

Given the consistent action of supersymmetry on the $L_\infty$ algebra of the theory, one naturally expects an analogous  action on the $C_\infty$ algebra upon color-stripping. This is described by a set of $n-$linear maps $\rho_n(\epsilon)$
\begin{equation}
    \rho_n(\epsilon):\;\cK_{\rm SYM}^{\otimes\,n} \longrightarrow \cK_{\rm SYM}\;,\quad |\rho_n(\epsilon)|=1-n\;.
\end{equation}
Similarly to the products $m_n$, they are obtained from the $\Sigma_n(\epsilon)$ maps acting on the $L_\infty$ algebra via
\begin{subequations}\begin{align}
    \Sigma_1(\epsilon\,|\,x) &= \rho_1(\epsilon\,|\,u^a)\otimes t_a\;, \\[1mm]
    \Sigma_2(\epsilon\,|\,x_1,x_2) &= (-1)^{x_1}f^a{}_{bc}\,\rho_2(\epsilon\,|\,u_1^b,u_2^c)\otimes t_a \,.
\end{align}\label{Eq:SUSY_Linf_Cinf_relations}\end{subequations}
From this definition it follows that the $\rho_n(\epsilon)$ obey similar compatibility conditions with the $m_n$. To the order we are interested in, these are given by
\begin{subequations}\label{Prop:SYM_GlobalSUSY_stripped}\begin{align}
    &m_1\rho_1(\epsilon\,|\,u) = \rho_1(\epsilon\,|\,m_1u)\;, \\[1mm]
    &m_1\rho_2(\epsilon\,|\,u,v)+\rho_2(\epsilon\,|\,m_1u,v)+(-1)^u\rho_2(\epsilon\,|\,u,m_1v)  \nonumber \\
    &\hspace{1.9cm} = \rho_1\big(\epsilon\,|\,m_2(u,v)\big)-m_2\big(\rho_1(\epsilon\,|\,u),v\big)-m_2\big(u,\rho_1(\epsilon\,|\,v)\big)\;.
\end{align}\end{subequations}
In the following, we will frequently use the abstract commutator notation $[m_1,\rho_1(\epsilon)]=0$ and $[m_1,\rho_2(\epsilon)]=[\rho_1(\epsilon),m_2]$ for the above and similar algebraic relations.

As we have discussed in the previous section, the $C_\infty$ algebra on $\cK_{\rm SYM}$ is only the surface structure  of the kinematic algebra BV$_\infty^\Box$. While the consistent action of supersymmetry on the $C_\infty$ sector is essentially guaranteed by the fact that the field theory is supersymmetric, requiring it to extend to the full BV$_\infty^\Box$ algebra is highly non-trivial. We encounter the first instance of this upon introducting the $b$ operator, since already at the linear level one needs to find a relation between $b$ and the linear map $\rho_1(\epsilon)$. Indeed, by making use of the properties \eqref{Prop:BV_inf_box} and \eqref{Prop:SYM_GlobalSUSY_stripped}, we can compute
\begin{equation}\begin{split}
    [m_1,[b,\rho_1(\epsilon)]] &= [[m_1,b],\rho_1(\epsilon)] - [b,\cancel{[m_1,\rho_1(\epsilon)]}] = [\Box,\rho_1(\epsilon)] = 0\;,
\end{split}\end{equation}
where in the last step we used the fact that $\Box$ commutes with any linear operator.
We thus find that $[b,\rho_1(\epsilon)]$ is $m_1$-closed. Modulo cohomological obstructions we expect it to be exact, which turns out to be the case:
\begin{equation}\label{Def:Theta_1_map}
    [b,\rho_1(\epsilon)]= [m_1,\Theta_{1}(\epsilon)]\;,  \quad |\Theta_{1}(\epsilon)| = -2.
\end{equation}
Using the map $\rho_1(\epsilon)$ inferred from the linearized transformations \eqref{Eq:SYM_SUSY_transf_linear} and \eqref{deltaequations}, the operator $\Theta_1(\epsilon)$ is found to be
\begin{equation}
    \Theta_1(\epsilon\,|\,\cE_{\rm F}) = -i\epsilon_A\widebar\cE^A\,\in\,K_0\;, \quad
    \Theta_1(\epsilon\,|\,\cN) = \bpm 0 \,,\; 0 \,,\; i\epsilon_AN \\[1mm] -i\epsilon_B\bar N^B\epm\,\in\,K_1\;,
\end{equation}
all the other components being zero.

While the relation \eqref{Def:Theta_1_map} per se defines a consistent action of supersymmetry on the kinematic algebra, it is problematic in view of the double copy. As we will discuss in the next section, the spectrum of the supergravity theory is constrained by the $b$ operator, and the relation \eqref{Def:Theta_1_map} would not preserve such constraint. To remedy this, we notice that supersymmetry of the field equations at the linearized level only requires $[m_1,\rho_1(\epsilon)]=0$. This does not fix the operator $\rho_1(\epsilon)$ completely, as one can shift it by an exact term
\begin{equation}\label{rhoshift}
    \rho_1(\epsilon)\;\longrightarrow\;\rho_1(\epsilon)+[m_1, H_1(\epsilon)]\;,   
\end{equation}
for any degree $-1$ operator $H_1(\epsilon)$. This is possible, intuitively, because the chain complex contains both auxiliary fields and dependent equations of motion, which make the action of $\rho_1(\epsilon)$ subject to arbitrary choices. In particular, if $\Theta_1(\epsilon)$ in \eqref{Def:Theta_1_map} is of the form $\Theta_1(\epsilon)=[b,H_1(\epsilon)]$ for some $H_1(\epsilon)$, one can choose a representative $\rho_1(\epsilon)$ in the class \eqref{rhoshift} that commutes with $b$.
Such a choice is indeed possible and we give in the following the explicit form of a $\rho_1(\epsilon)$ obeying
\begin{equation}
[m_1,\rho_1(\epsilon)]=0\;,\quad [b,\rho_1(\epsilon)]=0   \;. 
\end{equation}

To present the action of $\rho_1(\epsilon)$ on the color-stripped fields, we split them into bosons and fermions:
\begin{equation}
    \cA_{\rm B}\defeq \bpm A_\mu \,,\,\phi_{AB} \\[1mm] \varphi \epm \;,\quad
    \cA_{\rm F}\defeq \bpm \chi_A \,,\, \bar\chi^A \\[1mm]\varnothing\epm \;,   
\end{equation}
where we included both chiralities for the fermions. Here and in the following we use the empty set symbol $\varnothing$ to remind the reader that the fermionic fields belong to the upper row of the diagram \eqref{Fermibdiagram}, with respect to the $\mathbb{Z}_2$ grading implemented by the $b$ operator. The action of $\rho_1(\epsilon)$ on fields reads
\begin{subequations}\begin{align}
    % On bosons
    \rho_1(\epsilon\,|\,\cA_{\rm B}) &= \bpm 2\,\epsilon_A\sigma^{\mu\nu}\del_\mu A_\nu \;,\; 2\,\epsilon_B\sigma^\mu\del_\mu\phi^{AB}\\[1mm]\varnothing\epm\;, \\[3mm]
    % On fermions
    \rho_1(\epsilon\,|\,\cA_{\rm F}) &= \bpm \epsilon_A\sigma_\mu\widebar\chi^A \;,\; 2i\,\epsilon_{[A}\chi_{B]} \\[1mm]
    0 \epm\;.
\end{align}\end{subequations}
In particular, let us mention that the only difference with \eqref{Eq:SYM_SUSY_transf_linear} is that the auxiliary field $\varphi$ now does not transform.
In the same way, grouping the equations as
\begin{equation}
    \cE_{\rm B}\defeq \bpm E \\[1mm] E_\mu \,,\, E_{AB} \epm \;,\quad
    \cE_{\rm F}\defeq \bpm \cE_A \,,\, \bar\cE^A \\[1mm] E_A \,,\, \bar E^A \epm \;,   
\end{equation}
the action of $\rho_1(\epsilon\,|\,\cE)$ is given by
\begin{subequations}\begin{align}
    % On bosons
    \rho_1(\epsilon\,|\,\cE_{\rm B}) &= \bpm (E_\mu-\del_\mu E)\bar\sigma^\mu\epsilon_A \;,\; 2i\,E^{AB}\epsilon_B \\[2mm]
    -2\,\del_\mu E_\nu\sigma^{\mu\nu}\epsilon_A \;,\; 2\,\widebar\sigma^\mu\del_\mu E^{AB}\epsilon_B
    \epm\;, \\[3mm]
    % On fermions
    \rho_1(\epsilon\,|\,\cE_{\rm F}) &= \bpm -i\, \epsilon_A\widebar\cE^A \\[2mm]
    \epsilon_A\sigma_\mu\widebar E^A \;,\; -2i\,\epsilon_{[A}E_{B]}\epm\;,
\end{align}\end{subequations}
which differs from \eqref{deltaequations}. Similarly, the action on the Noether identities is
\begin{equation}
    \rho_1(\epsilon\,|\,\cN)
    = \bpm -i\, \epsilon_A N \;,\; 0 \\[2mm]
    -i\, \epsilon_A\widebar N^A \;,\; -\bar\sigma^\mu\epsilon_A\del_\mu N \;,\; 0\epm \;,
\end{equation}
and $\rho_1(\epsilon\,|\,\Lambda)=0$, since there are no gauge parameters into which $\Lambda$ can be rotated by supersymmetry.

As we have discussed, the linear map $\rho_1(\epsilon)$ displayed above is related by an exact shift \eqref{rhoshift} to the one generating the standard linear transformations \eqref{Eq:SYM_SUSY_transf_linear} and \eqref{deltaequations}. Accordingly, to preserve the relation $[m_1,\rho_2(\epsilon)]=[\rho_1(\epsilon),m_2]$, the bilinear map $\rho_2(\epsilon)$ has to be shifted, namely
\begin{subequations}\begin{align}
    \rho_1(\epsilon) \;&\longrightarrow\; \rho_1(\epsilon)+[m_1, H_1(\epsilon)]\;,\\[1mm]
    \rho_2(\epsilon) \;&\longrightarrow\; \rho_2(\epsilon)+[H_1(\epsilon),m_2]\;.
\end{align}\end{subequations}

%%%%%%%%%%%%%%%%%%%%%%%%%%%%%%%%%%
%%%%%%%%%%%%%%%%%%%%%%%%%%%%%%%%
%% BEGIN SECTION 3 %%%%%%%%%%%%%%%
%%%%%%%%%%%%%%%%%%%%%%%%%%%%%%%%
%%%%%%%%%%%%%%%%%%%%%%%%%%%%%%%%%%
\section{Double Copy of $\calN=4$ Super Yang-Mills}
\label{Sec:Double_Copy}

In this section we outline the general prescription for the double copy developed in \cite{Bonezzi:2022yuh,Bonezzi:2022bse}. It will be then applied to the special  case of $\calN=4$ super Yang-Mills theory to derive a \emph{double field theory} (DFT) version of $\calN=8$ supergravity. To this end, we will also propose a prescription for the double copy of the global supersymmetry, which will be performed in parallel to the double copy of the dynamical chain complex. We argue  that the resulting theory has to be the expected $\calN=8$ supergravity.

%%%%%%%%%%%%%%%%%%%%%%%%%%%%%%%%%
%%%%%%%%%%%%%%%%%%%%%%%%%%%%%%%%
%%%%%%%%%%%%%%%%%%%%%%%%%%%%%%%%%%
\subsection{General Construction}
\label{Subsec:General_DC}

The double copy procedure is a mathematical prescription that allows one to construct the scattering amplitudes of a gravity theory from the ones of a gauge theory \cite{Bern:2008qj}. From an off-shell point of view, such prescription is still not thoroughly established, leaving a gap in our understanding of the double copy as a relation between perturbative field theories. This is the reason why it is advantageous to treat both the gauge and gravity theories in the language of homotopy algebras, which has shown to be a fruitful framework to uncover deep algebraic structures, such as the kinematic BV$^\Box_\infty$ algebra.

Let us consider from the very beginning the example of $\calN=4$ super Yang-Mills theory, bearing in mind that the procedure that we are going to describe is more general and applies to any other gauge theory, as long as all fields are valued in the adjoint representation of the gauge group. As we have already seen in section \ref{Subsec:Linf_Form}, $\calN=4$ super Yang-Mills is described by the $L_\infty$ algebra $(\calX_\mathrm{SYM},B_n)$. This can be factorized as the tensor product of the kinematic algebra $\calK$ with the color Lie algebra $\mathfrak{g}$
\begin{equation}
    \calX_{\mathrm{SYM}} = \calK\otimes\mathfrak{g}\;.
\end{equation}
The first step towards the construction of a double copy supergravity theory is to strip the color Lie algebra off the gauge theory. This operation corresponds to what we have done in section \ref{Subsec:Kin_Alg} and leaves us with the kinematic algebra. To the order we are interested in, this consists of the graded vector space $\cK$, equipped with the $C_\infty$ products $\{m_n\}$ and the $b$ operator discussed in the previous section. We then consider a second copy $(\widetilde\calK,\tilde m_n,\tilde b)$ of such kinematic algebra and take the tensor product $\calK\otimes\widetilde{\calK}$. Since the elements of $\cK$ are local fields, the fields in $\cK\otimes\widetilde\cK$ are defined on a doubled spacetime with coordinates $(x^\mu,\tilde x^{\tilde\mu})$. As shown in \cite{Bonezzi:2022bse,Bonezzi:2023lkx}, upon restricting the functional dependence by the section constraint\footnote{In double field theory \cite{Hull:2009mi} this is the so-called strong constraint, which is related to the level matching of closed string theory.} $\Box=\widetilde\Box$, the space $\cK\otimes\widetilde\cK$ carries an $L_\infty$ algebra structure. This is the homotopy algebra underlying a double field theory version of $\calN=8$ supergravity, as it will be shown in the following. In a similar fashion to closed string field theory, the space $\cK\otimes\widetilde\cK$ contains twice as many elements compared to the DFT complex of supergravity \cite{Zwiebach:1992ie,Hull:2009mi}. The latter is given, besides the section constraint, by the linear subspace of elements $\Omega\in\cK\otimes\widetilde\cK$ obeying $b^-\Omega=0$, where
\begin{equation}
b^-\defeq \tfrac12\,\big(b\otimes\tilde\1-\1\otimes\tilde b\big)\;,\quad (b^-)^2=0\;.
    \label{Eq:b-constraint}
\end{equation}
Since the $b$ operator acts as a degree shift, the $b^-$ constraint projects out half of the elements of $\cK\otimes\widetilde\cK$, but does not impose any constraint on the remaining fields and parameters \cite{Bonezzi:2022yuh}.
Schematically, this process reads:
\begin{equation*}
    \left.\begin{array}{cc}
        \calX_{\mathrm{SYM}} = \calK\otimes\mathfrak{g} \quad\xrightarrow{\text{color-strip}}\quad \calK \; \\
        \\
        \widetilde\calX_{\mathrm{SYM}} = \widetilde\calK\otimes\widetilde{\mathfrak{g}} \quad\xrightarrow{\text{color-strip}}\quad \widetilde\calK \;
    \end{array}\right\}
    \quad\xrightarrow[\text{product}]{\text{tensor}}\quad %\calX_{\mathrm{SUGRA}}
    \calX_{\mathrm{SDFT}}
    \defeq \big(\calK\otimes\widetilde{\calK}\big)\big\rvert_{\rm section}\;,
\end{equation*}
where by $\big(\calK\otimes\widetilde{\calK}\big)\big\rvert_{\rm section}$ we denote the restriction to elements $\Omega$ of the tensor product obeying both $b^-\Omega=0$ and the section constraint $\Box=\widetilde\Box$.

At this point, one can sketch the structure of the graded vector space $\calX_{\mathrm{SDFT}}$ resulting from the double copy. This can be achieved by taking the tensor product of the elements of the chain complexes $\calK$ and $\widetilde\calK$ in different degrees. Accounting for the $b^-$ constraint this yields the following sets of elements
\begin{center}\begin{tabular}{ll}
    $\,\mathbbg{t} \defeq (\Uplambda\otimes\widetilde\Uplambda)$ & gauge for gauge parameters \\
    $\bbLambda \defeq (\Uplambda\otimes\widetilde\calA) \oplus (\calA\otimes\widetilde\Uplambda)$ & gauge parameters \\
    $\bbH \!\defeq (\calA\otimes\widetilde\calA) \oplus (\Uplambda\otimes\widetilde\calE\,) \oplus (\,\calE\otimes\widetilde\Uplambda)$ & fields \\
    $\bbE \defeq (\calA\otimes\widetilde\calE\,) \oplus (\,\calE\otimes\widetilde\calA) \oplus (\Uplambda\otimes\!\widetilde\calN) \oplus (\calN\otimes\widetilde\Uplambda)$ & equations of motion \\
    $\bbN \defeq (\,\calE\otimes\widetilde\calE\,) \oplus (\calA\otimes\!\widetilde\calN) \oplus (\calN\!\otimes\widetilde\calA)$ & Noether identities \\
    $\bbR \defeq (\,\calE\otimes\!\widetilde\calN) \oplus (\calN\!\otimes\widetilde\calE\,)$ & Noether for Noether identities \\
    \quad\quad\,\dots & \dots
\end{tabular}\end{center}
which can be arranged in a new $L_\infty$ chain complex
\begin{equation}
\label{CC:SuGra}
   \begin{tikzcd}[row sep=0pt, column sep={2.1cm,between origins}]
        X_{-2} \arrow[r,"\bbB_1"] &X_{-1} \arrow[r,"\bbB_1"] & X_{0} \arrow[r,"\bbB_1"]  &X_{1}\arrow[r,"\bbB_1"] &X_{2}\arrow[r,"\bbB_1"] &X_{3} \arrow[r,"\bbB_1"] & \cdots \\
        %%%
        \mathbbg{t} & \bbLambda & \mathbb{H} & \bbE & \bbN  & \mathbb{R} & \cdots
   \end{tikzcd}\quad.
\end{equation}
Notice that, given the semi-infinite fermionic complex of the gauge theory, $\calX_{\mathrm{SDFT}}$ is also semi-infinite. The spaces in degree higher than $3$ describe an infinite tower of trivial Noether identities, which are generated by dependent fermionic equations as in the description of super Yang-Mills theory of the previous section.

Given the $C_\infty$ products $m_1$ and $m_2$, together with the $b$ operator, the first $L_\infty$ brackets $\bbB_n$ of the double copy are given by \cite{Bonezzi:2022yuh,Bonezzi:2022bse}
\begin{subequations}\begin{align}
    \bbB_1 &\defeq m_1\otimes\tilde\1 + \1\otimes\tilde m_1 \label{Def:DC_B1}\;, \\[1mm]
    \bbB_2 &\defeq -\tfrac12 \,b^-(m_2\otimes\tilde m_2) \label{Def:DC_B2}\;,
\end{align}\end{subequations}
which will suffice for the purpose of this paper. 
Notice that, thanks to the section constraint and $(b^-)^2=0$, $b^-$ commutes with the differential $\bbB_1$ and annihilates $\bbB_2$
\begin{equation}\label{bminusconsistency}
[b^-,\bbB_1]=0\;,\quad b^-\bbB_2=0  \;,  
\end{equation}
implying that the restriction to ker$(b^-)$ is preserved and thus consistent.
By construction, the brackets above obey $\bbB_1^2=0$ and $[\bbB_1,\bbB_2]=0$, which ensure
that the double copy field equations
\begin{equation}
\bbB_1(\bbH)+\frac12\,\bbB_2(\bbH,\bbH)+\cdots=0    \;,
\end{equation}
are gauge covariant under $\delta_{\bbLambda}\bbH=\bbB_1(\bbLambda)+\bbB_2(\bbLambda,\bbH)+\cdots$ to first nonlinear order, where dots denote higher orders in the field $\bbH$. 

At this stage, the `super DFT' described by $(\cX_{\rm SDFT},\bbB_n)$ has fields that still depend, at least formally, on doubled coordinates and form representations of the doubled Lorentz group $\mathrm{Spin}(1,3)\times\widetilde{\mathrm{Spin}}(1,3)$. Moreover, due to the doubled $R$-symmetry, the spectrum is organized in multiplets of $SU(4)\times SU(4)$.
In order to recover standard supergravity in four dimensions, one can solve the section constraint by identifying the coordinates $x^\mu \equiv \tilde x^{\tmu}$, as well as the spinor indices $\alpha\equiv\tilde\alpha$, $\dot\alpha\equiv\dot{\tilde\alpha}$. After this identification, we will denote the resulting graded vector space by $\cX_{\rm SUGRA}$. The identification of coordinates and tangent spaces clearly breaks the doubled Lorentz group to the diagonal subgroup
\begin{equation}
    \mathrm{Spin}(1,3)\times\widetilde{\mathrm{Spin}}(1,3) \to \mathrm{Spin}(1,3)\;,
\end{equation}
which is the usual Lorentz group in four dimensions. Importantly, in doing so one does not identify the two copies of $SU(4)$ indices. This, as we shall see, is instrumental for enhancing 
the $R$-symmetry to $SU(8)$. 
%oh supersymmetry to $\cN=8$.

%%%%%%%%%%%%%%%%%%%%%%%%%%%%%%%%%
%%%%%%%%%%%%%%%%%%%%%%%%%%%%%%%%
%%%%%%%%%%%%%%%%%%%%%%%%%%%%%%%%%%
\subsection{Double Copy of  Global Supersymmetry}
\label{Subsec:DC_Global_SUSY}

In the previous section we have formulated the global $\cN=4$ supersymmetry of super Yang-Mills theory as an action on the kinematic algebra $\cK$, via the multilinear maps $\rho_n(\epsilon)$. In particular, we have established that one can define the linear map $\rho_1(\epsilon)$ so that it commutes with the $b$ operator. We summarize here the properties of the $\rho_n(\epsilon)$ maps for later convenience:
\begin{subequations}\label{Recaprhon}\begin{align}
    [m_1,\rho_1(\epsilon)] &=0\;, \label{Recaprhon1}\\[1mm]
    [b,\rho_1(\epsilon)] &=0\;, \label{Recaprhon2}\\[1mm]
    [m_1,\rho_2(\epsilon)] &=[\rho_1(\epsilon),m_2]\;, \label{Recaprhon3}
\end{align}\end{subequations}
where we recall  that the graded commutator between a linear operator and a bilinear product is defined as the failure of the operator to be a derivation of the product.

Given this structure on $\cK$, it is natural to ask how supersymmetry acts on the double copy. To this end, it is convenient to define a doubled global supersymmetry parameter
\begin{equation}\label{Globalepsilon8}
    \epsilon_I \defeq (\epsilon_A, \tilde\epsilon_{\tilde A}), \quad I=1,\dots,8 \;,
\end{equation}
together with the right-handed counterpart $\bar\epsilon^I$.
Similar to the differential $\bbB_1$, in the double copy theory we define the linear supersymmetry map
\begin{equation}\label{DCSigma1}
    \bbSigma_1(\epsilon) \defeq \rho_1(\epsilon)\otimes\tilde{\1} + \1\otimes\tilde{\rho}_1(\tilde\epsilon)\;,
\end{equation}
where the parameter on the left-hand side is the %oh doubled 
$\epsilon_I$ defined in (\ref{Globalepsilon8}). From this definition and \eqref{Def:DC_B1} one proves that $\bbB_1$ commutes with $\bbSigma_1(\epsilon)$, i.e.
\begin{equation}
    \bbB_1\circ\bbSigma_1(\epsilon) = \bbSigma_1(\epsilon)\circ\bbB_1\;,\quad{\rm or}\quad [\bbB_1,\bbSigma_1(\epsilon)]=0\;.
    \label{Prop:SUGRA_GlobalSUSY1}
\end{equation}
Using the relation \eqref{Recaprhon1} and recalling the degree $|\rho_1(\epsilon)|=0$, we have
\begin{equation}\begin{split}
\bbSigma_1(\epsilon)\bbB_1 &= \big(\rho_1(\epsilon)\otimes\tilde\1 + \1\otimes\tilde\rho_1(\tilde\epsilon)\big)(m_1\otimes\tilde\1 + \1\otimes\tilde m_1) \\
&= \rho_1(\epsilon)\,m_1\otimes\tilde{\1} + m_1\otimes\tilde\rho_1(\tilde\epsilon) + \rho_1(\epsilon)\otimes\tilde m_1 + \1\otimes\tilde\rho_1(\tilde\epsilon)\,\tilde m_1 \\
&= m_1\rho_1(\epsilon)\otimes\tilde{\1} + m_1\otimes\tilde\rho_1(\tilde\epsilon) + \rho_1(\epsilon)\otimes\tilde m_1 + \1\otimes\tilde m_1\tilde\rho_1(\tilde\epsilon) \\
&=(m_1\otimes\tilde\1 + \1\otimes\tilde m_1)\big(\rho_1(\epsilon)\otimes\tilde\1 + \1\otimes\tilde\rho_1(\tilde\epsilon)\big)\\
&= \bbB_1\bbSigma_1(\epsilon) \;,
\end{split}\end{equation}
which is \eqref{Prop:SUGRA_GlobalSUSY1}.
Given that $[b,\rho_1(\epsilon)]=0$, in the same way one shows
\begin{equation}\label{bminusSigma}
[b^-,\bbSigma_1(\epsilon)]=0\;,    
\end{equation}
which ensures that supersymmetry preserves the $b^-$ constraint and is thus well-defined on $\cX_{\rm SDFT}$. This, together with \eqref{Prop:SUGRA_GlobalSUSY1}, implies that the linearized double copy is supersymmetric, in the sense that the free field equations $\bbB_1(\bbH)=0$ are covariant under the transformations $\delta_\epsilon\bbH=\bbSigma_1(\epsilon\,|\,\bbH)$.

In perturbation theory, the full supersymmetry transformation is defined as a power series in the field $\bbH$ via
\begin{equation}
\delta_\epsilon\bbH=\bbSigma_1(\epsilon\,|\,\bbH)+\frac12\,\bbSigma_2(\epsilon\,|\,\bbH,\bbH)+\cdots    \;,
\end{equation}
which define the $n$-linear maps $\bbSigma_n(\epsilon\,|\,\bbH,\ldots,\bbH)$. In analogy with the SYM case \eqref{Prop:SYM_GlobalSUSY2}, covariance under supersymmetry to next order requires to find a $\bbSigma_2(\epsilon)$ obeying
\begin{equation}\label{Prop:SUGRA_GlobalSUSY2}
[\bbB_1,\bbSigma_2(\epsilon)]=[\bbSigma_1(\epsilon),\bbB_2] \;.   
\end{equation}
Such a $\bbSigma_2(\epsilon)$ can be found in a constructive way, upon using \eqref{Recaprhon}, \eqref{bminusSigma} and the $C_\infty$ relation $[m_1,m_2]=0$, which states that $m_1$ is a derivation of the product $m_2$. Using the expression
\begin{equation}
    \bbB_2= -\tfrac12\, b^-\big(m_2\otimes\tilde m_2\big)\;,
\end{equation}
for the two-bracket of the double copy, we compute the commutator 
\begin{equation}
\begin{split}
[\bbSigma_1(\epsilon),\bbB_2]&=-\tfrac12\,b^-\,[\bbSigma_1(\epsilon),m_2\otimes\tilde m_2]\\
&=-\tfrac12\,b^-\Big\{[\rho_1(\epsilon),m_2]\otimes\tilde m_2+m_2\otimes[\tilde\rho_1(\tilde\epsilon),\tilde m_2]\Big\}\\
&=-\tfrac12\,b^-\Big\{[m_1,\rho_2(\epsilon)]\otimes\tilde m_2+m_2\otimes[\tilde m_1,\tilde\rho_2(\tilde\epsilon)]\Big\}\\
&=-\tfrac12\,b^-\Big\{\big[m_1\otimes\tilde\1,\rho_2(\epsilon)\otimes\tilde m_2\big]+\big[\1\otimes\tilde m_1,m_2\otimes\tilde\rho_2(\tilde\epsilon)\big]\Big\}\\
&=-\tfrac12\,b^-\Big\{\big[\bbB_1,\rho_2(\epsilon)\otimes\tilde m_2+m_2\otimes\tilde\rho_2(\tilde\epsilon)\big]\Big\}\;,
\end{split}    
\end{equation}
where, to get to the last line, we used again $[m_1,m_2]=0$, together with its tilde counterpart, to reconstruct $\bbB_1$ by adding the missing terms $\1\otimes\tilde m_1$ and $m_1\otimes\tilde\1$. Finally, since $\bbB_1$ commutes with $b^-$, we can extract a total commutator and prove \eqref{Prop:SUGRA_GlobalSUSY2}, with $\bbSigma_2(\epsilon)$ given by the double copy formula
\begin{equation}
    \bbSigma_2(\epsilon) \defeq \tfrac12\,b^-\big( \rho_2(\epsilon)\otimes\tilde m_2+m_2\otimes\tilde\rho_2(\tilde\epsilon)\big)\;.    
\end{equation}
The above result also implies that $b^-\bbSigma_2(\epsilon)=0$, which is necessary to preserve the $b^-$ constraint under supersymmetry to this order.

In this perturbative setup, the above analysis shows how the double copy inherits a doubled global symmetry with eight spinor parameters $\epsilon_I=(\epsilon_A,\tilde\epsilon_{\tilde A})$. This, per se, does not imply that the 
$\cN=8$ supersymmetry algebra is obeyed. However, in the next section we will verify the presence of $\cN=8$ supersymmetry by both making contact with the standard formulation of $\cN=8$ supergravity, and by showing that the fields of the double copy organize themselves in multiplets of the $R$-symmetry group $SU(8)$.

%%%%%%%%%%%%%%%%%%%%%%%%%%%%%%%%%
%%%%%%%%%%%%%%%%%%%%%%%%%%%%%%%%
%%%%%%%%%%%%%%%%%%%%%%%%%%%%%%%%%%
\subsection{Explicit Example: The Free Theory}
\label{Subsec:DC_Chain_Cmplx}
As an explicit illustration of the general procedure outlined so far, we present the free theory obtained from the double copy. We first discuss the set of gauge parameters and spectrum of fields and then give the free field equations $\bbB_1(\bbH)=0$ and gauge transformations $\delta_{\bbLambda}\bbH$. We conclude this section with the global supersymmetry transformations of the super DFT.

Let us start from the set of gauge parameters $\bbLambda\in X_{-1}$, referring to the chain complex $(\cX_{\rm SDFT},\bbB_1)$ in \eqref{CC:SuGra}. According to the general discussion, this space is given by 
\begin{equation}
X_{-1}=\big[(K_0\otimes\widetilde K_1)\oplus(K_1\otimes\widetilde K_0)\big]\big\rvert_{\rm section}\;,    
\end{equation}
where we recall that we have to restrict to elements annihilated by $b^-$, subject to the section constraint. The gauge parameter of the super DFT can be decomposed as
\begin{equation}\label{DCLambda}
    \bbLambda = \bpm \Lambda_{\mu},\tilde\Lambda_{\tilde\mu} \\[1mm]
    \eta\epm\oplus\bpm\Lambda_{AB},\tilde\Lambda_{\Atilde\Btilde} \\[1mm]
    \varnothing \epm \oplus
    \bpm \varepsilon_{\Atilde}, \tilde\varepsilon_A \\[1mm]
    \varnothing \epm\;,    
\end{equation}
where $\Lambda_{AB}$ and $\tilde\Lambda_{\Atilde\Btilde}$ are antisymmetric in the $SU(4)$ indices. It consists of three sectors: The first contains two vector parameters $\Lambda_\mu$ and $\tilde\Lambda_{\tilde\mu}$, associated with linearized diffeomorphisms and $B$-field gauge transformations, plus a St\"uckelberg parameter $\eta$ related to auxiliary fields (which also appears in the original string field theory derivation of double field theory \cite{Hull:2009mi}). The second sector consists of $6+6$ scalar parameters $\Lambda_{AB}$ and $\tilde\Lambda_{\Atilde\Btilde}$ for $U(1)$ gauge transformations of vector fields. Finally, we have $4+4$ fermionic gauge parameters $(\varepsilon_{\Atilde},\tilde\varepsilon_A)$, together with their complex conjugates of opposite chirality. In this regard, the reality conditions on $\cX_{\rm SDFT}$ are inherited from \eqref{Eq:Reality_Cond_Scalars} and \eqref{Eq:Reality_Cond_Spinors} of the single copy constituents.

We now come to describe the space of fields $X_0$, which similarly arises as
\begin{equation}
    X_{0}=\big[(K_1\otimes\widetilde K_1)\oplus(K_0\otimes\widetilde K_2)\oplus(K_2\otimes\widetilde K_0)\big]\big\rvert_{\rm section}\;.    
\end{equation}
Due to the number of components, it is convenient to split the field content into sectors reminiscent of closed string theory, namely $\bbH=\bbH_{\rm NS-NS}\oplus\bbH_{\rm NS-R}\oplus\bbH_{\rm R-R}$.
\begin{itemize}
\item {\bf NS-NS sector:} Here the field content is given by the one of the original double field theory \cite{Hull:2009mi}, which also arises from the double copy of bosonic Yang-Mills theory \cite{Bonezzi:2022yuh}, together with a number of vector fields and scalars as follows
\begin{equation}\label{DCField NSNS}
    \bbH_{\rm NS-NS} = \bpm e_{\mu\tnu}, e, \tilde e \\[1mm]
    f_\mu, \tilde f_{\tilde\mu} \epm \oplus
    \bpm A_{\mu\Atilde\Btilde}, \Atilde_{\tilde\mu AB} \\[1mm]
    \varphi_{\Atilde\Btilde},\tilde\varphi_{AB} \epm \oplus
    \bpm \Phi_{AB\Ctilde\Dtilde} \\[1mm]
    \varnothing \epm\;,
\end{equation}
where couples $AB$ and $\tilde A\tilde B$ of $SU(4)$ indices are always antisymmetric.
The first group consists of the tensor field $e_{\mu\tilde\nu}$, two scalars $(e,\tilde e)$ and two auxiliary vectors $(f_\mu,\tilde f_{\tilde\mu})$. Upon identifying the doubled coordinates and Lorentz groups, they describe the graviton, $B$-field and dilaton. The second group of fields in \eqref{DCField NSNS} displays $6+6$ vectors $(A_{\mu\Atilde\Btilde},\Atilde_{\tilde\mu AB})$, associated with the $6+6$ gauge parameters of \eqref{DCLambda}, together with $6+6$ auxiliary scalars $(\varphi_{\Atilde\Btilde},\tilde\varphi_{AB})$, analogous to the auxiliary $\varphi$ introduced in super Yang-Mills theory. Finally, $\Phi_{AB\Ctilde\Dtilde}$ are 36 matter scalars.
\item {\bf NS-R sector:} Here we find the fermions of the theory. They can be further split into two groups as
\begin{equation}
    \bbH_{\rm NS-R} = \bpm \psi_{\mu\Atilde}, \widebar\rho_{\Atilde}, \tilde\psi_{\tilde\mu A}, \tilde{\widebar\rho}_A \\[1mm]
    \varphi_{\Atilde},\tilde\varphi_A \epm \oplus
    \bpm \chi_{AB\Ctilde}, \tilde\chi_{\Atilde\Btilde C} \\[1mm] \varnothing\epm \;.
\end{equation}
The first group contains $4+4$ spinor-vectors $(\psi_{\mu\Atilde},\tilde\psi_{\tilde\mu A})$ associated with the fermionic gauge parameters of \eqref{DCLambda}, $4+4$ spinors $(\widebar\rho_{\Atilde},\tilde{\widebar\rho}_A)$ and $4+4$ auxiliary spinors $(\varphi_{\Atilde},\tilde\varphi_A)$. Upon taking the supergravity solution of the section constraint, they give rise to $8$ spin 3/2 gravitini and $8$ spin 1/2 dilatini. The second group consists instead of $24+24$ matter spinors $(\chi_{AB\Ctilde}, \tilde\chi_{\Atilde\Btilde C})$. Let us mention that all spinors above, except for $(\widebar\rho_{\Atilde},\tilde{\widebar\rho}_A)$, are left-handed, where prior to identifying the two copies of $\mathrm{Spin}(1,3)$ we mean left-handed in both spinor spaces, while $(\widebar\rho_{\Atilde},\tilde{\widebar\rho}_A)$ are right-handed. For brevity, we omit to write all the opposite chiralities obtained by complex conjugation.
Note that the analogous assignment of spinor representations under the doubled Lorentz group has also been found in ${\cal N}=1$ double field theory 
\cite{Coimbra:2011nw,Hohm:2011nu,Jeon:2011sq}. 
\item {\bf R-R sector:} The last sector of the theory is described by
\begin{equation}
    \bbH_{\rm R-R}=\bpm F_{A\Btilde},F_A{}^{\Btilde} \\[1mm] \varnothing \epm \;,   
\end{equation}
which are $16+16$ bispinors. Since at the moment we have two copies of Spin(1,3), it is worth writing explicitly the type of spinor indices carried by these fields, together with their complex conjugates:
\begin{equation}
    \big(F_{A\tilde B}\big)_{\alpha\tilde\beta}\;\stackrel{*}{\longrightarrow}\;\big(\widebar{F}^{A\tilde B}\big)_{\dot\alpha\dot{\tilde\beta}}\;, \qquad
    \big(F_{A}{}^{\tilde B}\big)_{\alpha\dot{\tilde\beta}} \;\stackrel{*}{\longrightarrow}\;\big(\widebar{F}^A{}_{\tilde B}\big)_{\dot\alpha{\tilde\beta}}\;.
\end{equation}
Let us stress that, prior to identifying the two spinor spaces, these fields cannot be written in terms of the familiar Ramond-Ramond forms.
\end{itemize}

Having described the field content of the double copy, we now present the free field equations $\bbB_1(\bbH)=0$, which are gauge invariant under $\delta_{\bbLambda}\bbH=\bbB_1(\bbLambda)$. As we have done for the field content, we give them separately for the three sectors of the theory, starting with the NS-NS sector
\begin{subequations}\label{DCEomNSNS}\begin{align}
    \Box e_{\mu\tnu} + \tilde\del_{\tilde\nu}f_\mu - \del_\mu\tilde f_{\tilde\nu} &= 0\;, & 
    & \\[1mm]
    \del_\mu\tilde e - \tilde\del^{\tilde\nu}e_{\mu\tilde\nu} - f_\mu &= 0\;, & 
    \tilde\del_{\tilde\nu}e + \del^\mu e_{\mu\tilde\nu} - \tilde f_{\tilde\nu} &= 0\;, \\[1mm]
    \Box e - \del^\mu f_\mu &= 0\;, & 
    \Box\tilde e - \tilde\del^{\tilde\mu}\tilde f_{\tilde\mu} &= 0\;, \\[1mm]
    \Box A_{\mu\Atilde\Btilde} - \del_\mu\varphi_{\Atilde\Btilde} &= 0\;, & 
    \Box\Atilde_{\tilde\mu AB} + \tilde\del_{\tilde\mu}\tilde\varphi_{AB} &= 0\;, \\[1mm]
    \del^\mu A_{\mu\Atilde\Btilde} - \varphi_{\Atilde\Btilde} &= 0\;, & 
    \tilde\del^{\tilde\mu}\Atilde_{\tilde\mu AB} + \tilde\varphi_{AB} &= 0\;, \\[1mm]
    \Box\Phi_{AB\Ctilde\Dtilde} &= 0\;. & &
\end{align}\end{subequations}
One can see that, as we have anticipated, the vectors $f_\mu$ and $\tilde f_{\tilde\mu}$ as well as the scalars $\varphi_{\Atilde\Btilde}$ and $\tilde\varphi_{AB}$ are auxiliary fields that can be eliminated by their algebraic equations. The above field equations are invariant under the gauge transformations
\begin{subequations}\label{DCgaugeNSNS}\begin{align}
    \delta e_{\mu\tnu} &= \del_\mu\tilde\Lambda_{\tilde\nu} - \tilde\del_{\tilde\nu}\Lambda_\mu\;,\ &
    & \\[1mm]
    \delta e &= \del^\mu\Lambda_\mu - \eta\;, &
    \delta\tilde e &= \tilde\del^{\tilde\mu}\tilde\Lambda_{\tilde\mu} - \eta\;, \\[1mm]
    \delta f_\mu &= \Box\Lambda_\mu - \del_\mu\eta\;, &
    \delta\tilde f_{\tilde\mu} &= \Box\tilde\Lambda_{\tilde\mu} - \tilde\del_{\tilde\mu}\eta\;, \\[1mm]
    \delta A_{\mu\Atilde\Btilde} &= \del_\mu\tilde\Lambda_{\Atilde\Btilde}\;, &
    \delta\Atilde_{\tilde\mu AB} &= - \tilde\del_{\tilde\mu}\Lambda_{AB}\;, \\[1mm]
    \delta\varphi_{\Atilde\Btilde} &= \Box\tilde\Lambda_{\Atilde\Btilde}\;, &
    \delta\tilde\varphi_{AB} &= \Box\Lambda_{AB}\;, \\[1mm]
    \delta\Phi_{AB\Ctilde\Dtilde} &= 0\;. & &
\end{align}\end{subequations}
The gauge symmetries associated with $(\Lambda_\mu,\tilde\Lambda_{\tilde\mu}, \eta)$ are reducible, in that parameters of the form $\Lambda^{\rm triv.}_\mu=\del_\mu\tau$, $\tilde\Lambda^{\rm triv.}_{\tilde\mu}=\tilde\del_{\tilde\mu}\tau$ and $\eta^{\rm triv.}=\Box\tau$ do not generate any transformation on the fields. This is related, upon taking the supergravity solution of the section constraint, to the presence of a two-form in $e_{\mu\tilde\nu}$.

Moving to the NS-R sector, we obtain the following field equations for the fermions:
\begin{subequations}\label{DCEomNSR}\begin{align}
    i\,\widebar\sigma\!\cdot\!\tilde\del\psi_{\mu\Atilde} - \del_\mu\bar\rho_{\Atilde} &= 0\;, &
    i\,\widebar\sigma\!\cdot\!\del\tilde\psi_{\tilde\mu A} - \tilde\del_{\tilde\mu}\tilde{\bar\rho}_A &= 0\;, \\[1mm]
    \del^\mu\psi_{\mu\Atilde} - \varphi_{\Atilde} &= 0\;, &
    \tilde\del^{\tilde\mu}\tilde\psi_{\tilde\mu A} - \tilde\varphi_A &= 0\;, \\[1mm]
    i\,\sigma\!\cdot\!\tilde\del\bar\rho_{\Atilde} + \varphi_{\Atilde} &= 0\;, &
    i\,\sigma\!\cdot\!\del\tilde{\bar\rho}_A + \tilde\varphi_A &= 0\;, \\[1mm]
    i\,\widebar\sigma\!\cdot\!\tilde\del\chi_{AB\Ctilde} &= 0\;, &
    i\,\widebar\sigma\!\cdot\!\del\tilde\chi_{\Atilde\Btilde C} &= 0\;.
\end{align}\end{subequations}
To clarify the notation, we do not put a tilde on the second copy of the sigma matrices and we distinguish between the two by the vector index they carry, e.g. $\sigma^\mu$ and $\sigma^{\tilde\mu}$.
In the same fashion as with the super Yang-Mills fermions, the above equations are accompanied by dependent ones, which we do not write, given by acting on \eqref{DCEomNSR} with the massless Dirac operator. This complies with the infinite tower of trivial Noether identities contained in $\cX_{\rm SDFT}$. The fermionic gauge symmetries are given by
\begin{subequations}\label{DCgaugeNSR}\begin{align}
    \delta\psi_{\mu\Atilde} &= \del_\mu\varepsilon_{\Atilde}\;, &
    \delta\tilde\psi_{\tilde\mu A} &= \tilde\del_{\tilde\mu}\tilde\varepsilon_A\;, \\[1mm]
    \delta\bar\rho_{\Atilde} &= i\,\widebar\sigma\!\cdot\!\tilde\del\,\varepsilon_{\Atilde}\;, &
    \delta\tilde{\bar\rho}_A &= i\,\widebar\sigma\!\cdot\!\del\,\tilde\varepsilon_A\;, \\[1mm]
    \delta\varphi_{\Atilde} &= \Box\,\varepsilon_{\Atilde}\;, &
    \delta\tilde\varphi_A &= \Box\,\tilde\varepsilon_A\;, \\[1mm]
    \delta\chi_{AB\tilde C} &=0\;, &
    \delta\tilde\chi_{\tilde A\tilde B C} &=0\;,
\end{align}\end{subequations}
with the expected transformations for the spin 3/2 fermions.

Finally, the Ramond-Ramond bispinors are gauge invariant and obey the massless Dirac equation in both spinor indices, i.e.
\begin{subequations}\label{DCEomRR}\begin{align}
    i\,\widebar\sigma\!\cdot\!\del F_{A\tilde B} &= i\,\widebar\sigma\!\cdot\!\tilde\del F_{A\tilde B}=0\;, \\[1mm]
    i\,\widebar\sigma\!\cdot\!\del F_{A}{}^{\tilde B} &=i \,\sigma\!\cdot\!\tilde\del F_{A}{}^{\tilde B}=0\;.
\end{align}\end{subequations}

As anticipated, the super DFT emerging from the double copy procedure is invariant under global $\cN=8$ supersymmetry. In particular, under global supersymmetry the fields belonging to the NS-NS sector transform as
\begin{subequations}\label{DC_SUSY_transf_NS-NS}\begin{align}
    \delta_\epsilon e_{\mu\tnu} &= \epsilon_A\sigma_\mu\tilde{\widebar\psi}_\tnu{}^A + \tilde\epsilon_\Atilde\sigma_\tnu\widebar\psi_\mu{}^\Atilde \;, &
    & \\[1mm]
    \delta_\epsilon e &= -i\, \epsilon_A\tilde\rho^A \;, &
    \delta_\epsilon \tilde e &= -i\, \tilde\epsilon_\Atilde\rho^\Atilde \;,\\[1mm]
    \delta_\epsilon f_\mu &= \epsilon_A\sigma_\mu\tilde{\widebar\varphi}^A \;, &
    \delta_\epsilon \tilde f_{\tilde\mu} &= \tilde\epsilon_\Atilde\sigma_\tmu\widebar\varphi^\Atilde \;,\\[1mm]
    \delta_\epsilon A_{\mu\Atilde\Btilde} &= 2i\, \tilde\epsilon_{[\Atilde}\psi_{\mu\Btilde]} \;, &
    \delta_\epsilon \Atilde_{\tilde\mu AB} &= 2i\, \epsilon_{[A}\tilde\psi_{\tmu B]} \;,\\[1mm]
    \delta_\epsilon \varphi_{\Atilde\Btilde} &= 2i\, \tilde\epsilon_{[\Atilde}\varphi_{\Btilde]} \;, &
    \delta_\epsilon \tilde\varphi_{AB} &= 2i\, \epsilon_{[A}\tilde\varphi_{B]} \;,\\[1mm]
    \delta_\epsilon \Phi_{AB\Ctilde\Dtilde} &= 2i\, \epsilon_{[A}\chi_{B]\Ctilde\Dtilde} + 2i\, \epsilon_{[\Atilde}\chi_{\Btilde]CD} \;.
\end{align}\end{subequations}
The fields in the NS-R sector, instead, transform as
\begin{subequations}\begin{align}
    \delta_\epsilon \psi_{\mu\Atilde} &= 2\, \tilde\epsilon_\Atilde\sigma^{\tnu\trho}\tilde\del_\tnu e_{\trho\mu} + \epsilon_B\sigma_\mu F^B{}_\Atilde \;, &
    \delta_\epsilon \tilde\psi_{\tilde\mu A} &=  2\, \epsilon_A\sigma^{\nu\rho}\del_\nu e_{\rho\tmu} + \tilde\epsilon_\Btilde\sigma_\tmu F_A{}^\Btilde \;, \\[1mm]
    \delta_\epsilon \widebar\psi_\mu{}^\Atilde &= 2\, \tilde\epsilon_\Btilde\sigma^\tnu\tilde\del_\tnu A_\mu{}^{\Atilde\Btilde} + \epsilon_B\sigma_\mu F^{\Atilde B} \;, &
    \delta_\epsilon \tilde{\widebar\psi}_\tmu{}^A &= 2\, \epsilon_B\sigma^\nu\del_\nu \Atilde_\tmu{}^{AB} + \tilde\epsilon_\Btilde\sigma_\tmu F^{A\Btilde} \;, \\[1mm]
    \delta_\epsilon \bar\rho_{\Atilde} &= \tilde\epsilon_\Atilde\sigma^\tmu\tilde\del_\tmu\tilde e - \tilde\epsilon_\Atilde \sigma^\tmu \tilde f_\tmu \;, &
    \delta_\epsilon \tilde{\bar\rho}_A &= \epsilon_A\sigma^\mu\del_\mu e - \epsilon_A \sigma^\mu f_\mu \;, \\[1mm]
    \delta_\epsilon \rho^{\Atilde} &= 2i\, \varphi^{\Atilde\Btilde}\tilde\epsilon_\Btilde \;, &
    \delta_\epsilon \tilde{\rho}^A &= 2i\, \tilde\varphi^{AB}\epsilon_B \;, \\[1mm]
    \delta_\epsilon \varphi_{\Atilde} &= 2\, \tilde\epsilon_\Atilde\sigma^{\tmu\tnu}\del_\tmu\tilde f_\tnu \;, &
    \delta_\epsilon \tilde\varphi_A &= 2\, \epsilon_A\sigma^{\mu\nu}\del_\mu f_\nu \;, \\[1mm]
    \delta_\epsilon \widebar\varphi^{\Atilde} &= 2\, \tilde\epsilon_\Btilde\sigma^\tmu\tilde\del_\mu\varphi^{\Atilde\Btilde} \;, &
    \delta_\epsilon \tilde{\widebar\varphi}^A &= 2\, \epsilon_B\sigma^\mu\del_\mu\tilde\varphi^{AB} \;, \\[1mm]
    \delta_\epsilon \chi_{AB\tilde C}&= 2i\, \epsilon_{[A}F_{B]\Ctilde} \;, &
    \delta_\epsilon \tilde\chi_{\tilde A\tilde B C} &= 2i\, \tilde\epsilon_{[\Atilde}F_{\Btilde]C} \;, \\[1mm]
    \delta_\epsilon \chi^{A\Btilde\tilde C} &= 2\, \epsilon_{D}\sigma^\mu\del_\mu\Phi^{AD\Btilde\Ctilde} \;, &
    \delta_\epsilon \chi^{AB\tilde C} &= 2\, \tilde\epsilon_\Dtilde\sigma^\tmu\tilde\del_\tmu\Phi^{AB\Ctilde\Dtilde} \;, \\[1mm]
    \delta_\epsilon \chi^{\Atilde\Btilde}{}_C &= 2\, \epsilon_C\sigma^{\mu\nu}\del_\mu A_\nu{}^{\Atilde\Btilde} \;, &
    \delta_\epsilon \chi^{AB}{}_\Ctilde &= 2\, \tilde\epsilon_\Ctilde\sigma^{\tmu\tnu}\del_\tmu \Atilde_\tnu{}^{AB} \;, \\[1mm]
    \delta_\epsilon \chi_{AB}{}^\Ctilde &= 2i\, \epsilon_{[A}F_{B]}{}^{\Ctilde} \;, &
    \delta_\epsilon \chi_{\Atilde\Btilde}{}^C &= 2i\, \tilde\epsilon_{[\Atilde}F_{\Btilde]}{}^C \;.
\end{align}\end{subequations}
To conclude, the R-R sector is characterized by the following global transformations:
\begin{subequations}\label{DC_SUSY_transf_R-R}\begin{align}
    \delta_\epsilon F_{A\Btilde} &= 2\, \epsilon_A\sigma^{\mu\nu}\del_\mu\psi_{\nu\Btilde} + 2\, \tilde\epsilon_\Btilde\sigma^{\tmu\tnu}\del_\tmu\tilde\psi_{\tnu A} \;,\\[1mm]
    \delta_\epsilon F^{A\Btilde} &= 0 \;,\\[1mm]
    \delta_\epsilon F_A{}^\Btilde &= 2\, \epsilon_A\sigma^{\mu\nu}\del_\mu\widebar\psi_\nu{}^\Btilde + \epsilon^{\Btilde\Ctilde\tilde L\tilde M}\tilde\epsilon_\Ctilde\sigma^\tmu\del_\tmu\chi_{A\tilde L\tilde M} \;,\\[1mm]
    \delta_\epsilon F^A{}_\Btilde &= 2\, \tilde\epsilon_\Btilde\sigma^{\tmu\tnu}\del_\tmu\tilde{\widebar\psi}_\tnu{}^A + \epsilon^{ACLM}\epsilon_C\sigma^\mu\del_\mu\chi_{LM\Btilde} \;.
\end{align}\end{subequations}
Notice that the field-strengths contained in the R-R sector are naturally transformed into new field-strengths by supersymmetry.
Again, the global transformations with respect to the conjugate parameters $\bar\epsilon^A, \tilde{\bar\epsilon}^\Atilde$ are meant to be added to the above-listed ones.

In the next section we will take the supergravity solution of the section constraint and make contact with the standard formulation of $\cN=8$ supergravity at the linearized level.

%%%%%%%%%%%%%%%%%%%%%%%%%%%%%%%%%%
%%%%%%%%%%%%%%%%%%%%%%%%%%%%%%%%
%% BEGIN SECTION 4 %%%%%%%%%%%%%%%
%%%%%%%%%%%%%%%%%%%%%%%%%%%%%%%%
%%%%%%%%%%%%%%%%%%%%%%%%%%%%%%%%%%
\section{$\calN=8$ Supergravity}
\label{Sec:N=8_SUGRA_as_DC}

In this section we will first show that the field content arising from the double copy matches the one of $\calN=8$ supergravity, upon taking an explicit solution of the section constraint. With this solution, we will demonstrate that the double copy free field equations \eqref{DCEomNSNS}, \eqref{DCEomNSR} and \eqref{DCEomRR} are equivalent to the standard supergravity ones. We conclude by exhibiting how the fields can be arranged in representations of $SU(8)$, which is the $R$-symmetry group of $\cN=8$ supergravity.

%%%%%%%%%%%%%%%%%%%%%%%%%%%%%%%%%
%%%%%%%%%%%%%%%%%%%%%%%%%%%%%%%%
%%%%%%%%%%%%%%%%%%%%%%%%%%%%%%%%%%
\subsection{Spectrum of $\calN=8$ Supergravity}
\label{Subsec:Lin_N=8_SUGRA}
    %3.4 Linearised N=8 SuGra (RR sector + massive deformation, RS equation)
Let us start by analyzing in detail the spectrum obtained in the previous section. We first solve the section constraint by identifying coordinates $\tilde x^\tmu\equiv x^\mu$ and Lorentz groups $\widetilde{\mathrm{Spin}}(1,3)\equiv\mathrm{Spin}(1,3)$. In practice, this amounts to identifying derivatives $\tilde\del_{\tilde\mu}\equiv\del_\mu$, vector and spinor indices $\tilde\mu\equiv\mu$, $\tilde\alpha\equiv\alpha$ and $\dot{\tilde\alpha}\equiv\dot{\alpha}$, as well as the two copies of the Minkowski metric and sigma matrices: $\eta_{\tilde\mu\tilde\nu}\equiv\eta_{\mu\nu}$, $\sigma^{\tilde\mu}\equiv\sigma^\mu$. In order to analyze the spectrum, in the following we will further assume that all auxiliary fields have been eliminated by means of their equations of motion.

\subsubsection*{NS-NS Sector}
We start with the  tensor $e_{\mu\nu}$, which can now be split into its symmetric and antisymmetric parts
\begin{equation}\label{graviton defined}
e_{\mu\nu} = h_{\mu\nu} + B_{\mu\nu}\;,\quad h_{\mu\nu}\defeq e_{(\mu\nu)}\;,\quad B_{\mu\nu}
\defeq e_{[\mu\nu]}\;,
\end{equation}
$h_{\mu\nu}$ being the metric fluctuation and $B_{\mu\nu}$ being the antisymmetric rank-2 tensor, known in the string theory literature as \emph{$B$-field}. 
The vector gauge parameters can also be redefined as
\begin{equation}
\xi_\mu\defeq \tfrac12\,(\tilde\Lambda_\mu-\Lambda_\mu)\;,\quad\zeta_\mu\defeq \tfrac12\,(\tilde\Lambda_\mu+\Lambda_\mu) \;.   
\end{equation}
The gauge transformations of $h_{\mu\nu}$ and $B_{\mu\nu}$ are then the standard ones of a massless spin two field and a two-form gauge potential:
\begin{equation}
\delta h_{\mu\nu}=\del_\mu\xi_\nu+\del_\nu\xi_\mu\;,\quad \delta B_{\mu\nu}=\del_\mu\zeta_\nu-\del_\nu\zeta_\mu\;,  
\end{equation}
confirming that $\xi_\mu$ is the linearized diffeomorphism parameter. This also explains the reducibility mentioned previously, as the two-form has trivial parameters $\zeta_\mu^{\rm triv.}=\del_\mu\tau$.
In four dimensions, a two-form potential carries a single propagating degree of freedom and it can be dualized on-shell into a pseudo-scalar $a$ via the relation
\begin{equation}
    \del^\mu a = \frac{1}{3!}\epsilon^{\mu\nu\rho\sigma}\del_\nu B_{\rho\sigma}\,.
\end{equation}

Besides the tensor $e_{\mu\nu}$ and the auxiliary fields $f_\mu$ and $\tilde f_\mu$, the DFT subsector contains the two scalars $e$ and $\tilde e$. These can be combined to define a gauge invariant dilaton
\begin{equation}\label{dilaton defined}
\phi\defeq \tfrac12\,(\tilde e-e-h^\mu{}_\mu)\;,\quad\delta\phi=0\;,
\end{equation}
while the other combination is pure gauge: $\delta(\tilde e+e)=2\,\del^\mu\zeta_\mu-2\,\eta$.
As we shall see later, this combination
does not appear in the equations of motion and can be gauged away by a St\"uckelberg transformation with parameter $\eta$. Notice that the gauge invariant dilaton \eqref{dilaton defined} can only be defined after solving the section constraint, as there is no way of taking the trace of $e_{\mu\tilde\nu}$ without breaking the doubled Lorentz symmetry.

The NS-NS sector also contains the vector fields $A_{\mu AB}$ and $A_{\mu\Atilde\Btilde}$ with their standard gauge symmetry, together with
the gauge invariant scalars $\Phi_{AB\Ctilde\Dtilde}$.
Altogether, this part of the spectrum accounts for one graviton $h_{\mu\nu}$, 12 vectors $A_{\mu AB}\,, A_{\mu \tilde A\tilde B}$ and 38 scalars: the 36 $\Phi_{AB\tilde C\tilde D}$ plus the dilaton $\phi$ and the axion $a$ coming from the $B$-field.
Let us stress that this counting is for real fields. However, the reality conditions of the vectors and $\Phi_{AB\Ctilde\Dtilde}$ scalars are not straightforward and will be discussed in detail at the end of the section. 

\subsubsection*{NS-R Sector}
% % % Dilatini
Prior to solving the section constraint, one cannot identify the vector-spinors as spin 3/2 fermions. This is because their vector and spinor indices belong to the two different copies of the Lorentz group, e.g. $\psi_{\mu\tilde\alpha\,\tilde A}$. Upon identifying the two Lorentz groups, $\psi_{\mu A}$ and $\psi_{\mu \tilde A}$ are genuine spin 3/2 fermions, to be viewed as the eight gravitini of the supergravity theory. Having access to the $\sigma$-trace, one can define eight gauge invariant spin 1/2 dilatini via
\begin{equation}\label{Eq:Dilatino}    
\bar\lambda_A = \bar\rho_A -i\,\widebar\sigma^\mu\psi_{\mu A}\;,\quad
\bar\lambda_\Atilde = \bar\rho_\Atilde -i\,\widebar\sigma^\mu\psi_{\mu\Atilde}\;,
\end{equation}
together with the left-handed complex conjugates. The remaining 48 fermions $\chi_{A\tilde B\tilde C}$ and $\chi_{\tilde ABC}$ are already gauge invariant and sum up with the left-handed dilatini $\lambda^A$, $\lambda^{\tilde A}$ for a total of 56 Weyl spinors.

\subsubsection*{R-R Sector}
As we have seen in the previous section, the Ramond-Ramond sector consists of gauge invariant bispinors which, after identifying the Lorentz groups, have the following spinor structure:
\begin{equation}
\big(F_{A\tilde B}\big)_{\alpha\beta}\;,\quad \big(F_A{}^{\tilde B}\big)_{\alpha\dot\beta}\;,    
\end{equation}
together with their complex conjugates $\bar F^{A\tilde B}$ and $\bar F^A{}_{\tilde B}$. Using the Fierz identities
\begin{equation}\label{Eq:FierzI}
\xi_\alpha\widebar\zeta_{\dot\alpha} = \tfrac12\,\sigma^{\mu}_{\alpha\dot\alpha}\,(\xi\sigma_\mu\widebar\zeta)\;,\quad \xi_\alpha\zeta_\beta = \tfrac12\epsilon_{\alpha\beta}(\xi\zeta) + \tfrac12 \sigma^{\mu\nu}_{\alpha\beta} (\xi\sigma_{\mu\nu}\zeta)   \;, 
\end{equation}
one can express the bispinors as a combination of complex scalars $M_{A\tilde B}$, vectors $P_{\mu A}{}^{\tilde B}$ and self-dual two-forms $F^+\!\!\!\!\!{}_{\mu\nu A\tilde B}$ as  
\begin{equation}\label{RRFierzed}
F_A{}^{\tilde B}=\sigma^\mu P_{\mu A}{}^{\tilde B}\;, \quad F_{A\tilde B}=\epsilon\,M_{A\tilde B}+\tfrac12\,\sigma^{\mu\nu}F^+\!\!\!\!\!{}_{\mu\nu A\tilde B}  \;. 
\end{equation}
Similarly, the conjugate bispinors $\bar F^{A\tilde B}$ and $\bar F^A{}_{\tilde B}$ can be expanded in terms of the complex conjugate fields $\widebar M^{A\tilde B}$, $\widebar P_\mu^A{}_{\tilde B}$ and anti-self-dual two-forms $\bar F_{\mu\nu}^{-\,A\tilde B}$. These fields are gauge invariant and obey first-order equations of motion. The $P_{\mu A}{}^{\tilde B}$ are thus interpreted as field strengths for 16 complex scalars $\Phi_A{}^{\tilde B}$. Similarly, the two-forms are field strengths for 16 self-dual vectors $A^+\!\!\!{}_{\mu A\tilde B}$, with complex conjugate anti-self-dual $\bar A_{\mu}^{-\,A\tilde B}$. Finally, the scalars $M_{A\tilde B}$ are topological and carry no local degrees of freedom.
The reality conditions of all these fields will be discussed in detail at the end of this section.

\subsubsection*{Complete Physical Spectrum}

To summarize, taking care of the reality conditions the complete space of physical fields includes:
\begin{enumerate}
 \item[$\bullet$] \textbf{1 graviton:} $h_{\mu\nu}$ ($\times 1$);  
  \item[$\bullet$] \textbf{8 gravitini:} $\psi_{\mu A}$ ($\times 4$), $\psi_{\mu \Atilde}$ ($\times 4$);
\item[$\bullet$] \textbf{28 vectors:} $A_{\mu AB}$ ($\times 6$), $A_{\mu \Atilde\Btilde}$ ($\times 6$), $(A^+\!\!\!{}_{\mu A\tilde B},\bar A_{\mu}^{-\,A\tilde B})$ ($\times 16$);
  \item[$\bullet$] \textbf{56 Weyl spinors:} $\chi_{AB\Ctilde}$ ($\times 24$), $\chi_{\Atilde\Btilde C}$ ($\times 24$), $\lambda^A$ ($\times 4$), $\lambda^{\tilde A}$ ($\times 4$);   
    \item[$\bullet$] \textbf{70 real scalars:} $a$ ($\times 1$), $\phi$ ($\times 1$), $\Phi_{AB\Ctilde\Dtilde}$ ($\times 36$), $\Phi_A{}^{\Btilde}$ ($\times 16$ complex).
\end{enumerate}
This counting of fields matches the off-shell spectrum of $\calN=8$ supergravity. In the following we will provide further evidence that the double copy constructed here is a double field theory formulation of the standard $\calN=8$ theory. 

\subsection{Field Equations in the Supergravity Basis}

In order to show that the free dynamics of the double copy is equivalent to the standard supergravity one, we analyze  the field equations presented in section~\ref{Subsec:DC_Chain_Cmplx} in terms of the field basis described above.
We will focus on the gravitational, spin 3/2 and Ramond-Ramond sectors, which are subject to unconventional field equations. As we have mentioned before, we will assume that all auxiliary fields have been eliminated by using their equations of motion.

\subsubsection*{NS-NS Sector}

In the NS-NS sector, the equations \eqref{DCEomNSNS} for the vector and scalar fields are already in the standard form of Maxwell and massless Klein-Gordon equations:
\begin{equation}
    \del^\mu F_{\mu\nu AB}=\del^\mu F_{\mu\nu \tilde A\tilde B}=0\;,\quad\Box\Phi_{AB\tilde C\tilde D} =0\;,   
\end{equation}
written in terms of the abelian field strengths for $A_{\mu AB}$ and $A_{\mu \tilde A\tilde B}$. Using the definitions \eqref{graviton defined} and \eqref{dilaton defined} for the graviton, $B$-field and dilaton, the corresponding equations in \eqref{DCEomNSNS} read
\begin{subequations}\label{gravEom1}\begin{align}
    \Box h_{\mu\nu}-2\,\del_{(\mu}\del^\rho h_{\nu)\rho}+\del_\mu\del_\nu h+2\,\del_\mu\del_\nu\phi&=0\;,\\[1mm]
    2\,\Box\phi+\Box h-\del^\mu\del^\nu h_{\mu\nu}&=0\;,\\[1mm]
    \Box B_{\mu\nu}+2\,\del_{[\mu}\del^\rho B_{\nu]\rho}&=0\;,
\end{align}\end{subequations}
where $h\defeq h^\mu{}_\mu$ is the trace of the graviton field. As promised, only the gauge invariant combination of the scalars $e$ and $\tilde e$ appears in the field equations.  

The $B$-field is already decoupled and its equation of motion takes the standard form
\begin{equation}
\del^\rho H_{\rho\mu\nu}=0\;,\quad H_{\mu\nu\rho}\defeq 3\,\del_{[\mu}B_{\nu\rho]}\;,    
\end{equation}
in terms of the gauge invariant curvature $H_{\mu\nu\rho}$. In four dimensions, this can be dualized on-shell to a scalar field obeying the massless Klein-Gordon equation via
\begin{equation}
\frac{1}{3!}\,\epsilon_{\mu\nu\rho\sigma}\,H^{\nu\rho\sigma}=\del_\mu a\;,\quad\Box a=0\;.  
\end{equation}
The coupled graviton-dilaton system in \eqref{gravEom1} is equivalent to
\begin{equation}\label{gravEom2}
\Box h_{\mu\nu}-2\,\del_{(\mu}\del^\rho h_{\nu)\rho}+\del_\mu\del_\nu h+2\,\del_\mu\del_\nu\phi=0\;,\quad
\Box\phi=0\;,     
\end{equation}
which can be seen by
combining the trace of the spin two equation of motion with the scalar one.
This form of the field equation, with the dilaton linearly mixed with the graviton, signals that the spin two field $h_{\mu\nu}$ is the so-called string frame graviton. To decouple the spin two field from the scalar we perform a linearized Weyl transformation, thereby defining the Einstein frame graviton via
\begin{equation}
h_{\mu\nu}^{\rm E}\defeq h_{\mu\nu}+\eta_{\mu\nu}\,\phi\;. 
\end{equation}
The field equations are now completely decoupled and read
\begin{equation}
\Box h^{\rm E}_{\mu\nu}-2\,\del_{(\mu}\del^\rho h^{\rm E}_{\nu)\rho}+\del_\mu\del_\nu h^{\rm E}=0\;,\quad
\Box\phi=0\;,     
\quad\Box a=0\;.    
\end{equation}
The spin two equation is the standard one arising from the Fierz-Pauli lagrangian and is proportional to the linearized Ricci tensor $R^{\rm lin}_{\mu\nu}$ for the metric $g_{\mu\nu}=\eta_{\mu\nu}+h_{\mu\nu}^{\rm E}$.

\subsubsection*{NS-R Sector}

The matter fermions $\chi_{A\tilde B\tilde C}$ and $\chi_{\tilde A BC}$ are gauge invariant and obey the usual massless Dirac equation. We thus focus on the spin 3/2 subsector of \eqref{DCEomNSR}. Upon solving for the auxiliary spinors and introducing the gauge invariant dilatini \eqref{Eq:Dilatino}, the equations of motion take the form
\begin{subequations}\label{sugraEom1}\begin{align}
    i\,\widebar\sigma^\mu(\del_\mu\psi_{\nu A}-\del_\nu\psi_{\mu A})-\del_\nu\bar\lambda_A&=0\;,\\[1mm]
    \del^\mu\psi_{\mu A}-\sigma^\mu\widebar\sigma^\nu\del_\mu\psi_{\nu A}+i\,\sigma^\mu\del_\mu\bar\lambda_A&=0\;,
\end{align}\end{subequations}
together with an identical set of equations for $\psi_{\mu\tilde A}$ and $\bar\lambda_{\tilde A}$. Using the identity 
\begin{equation}\label{sigmasigmabar}
    \sigma^\mu\widebar\sigma^\nu = \eta^{\mu\nu}-2i\,\sigma^{\mu\nu} \;, 
\end{equation}
the equations \eqref{sugraEom1} can be written in terms of the gauge invariant fermionic field strength $\Psi_{\mu\nu A}\defeq \del_\mu\psi_{\nu A}-\del_\nu\psi_{\mu A}$ and simplify to
\begin{equation}\label{sugraEom2}
i\,\widebar\sigma^\mu\Psi_{\mu\nu A}-\del_\nu\bar\lambda_A=0\;,\quad
i\,\sigma^{\mu\nu}\Psi_{\mu\nu A}+i\,\sigma^\mu\del_\mu\bar\lambda_A=0\;.
\end{equation}

In a similar fashion as the spin two sector, combining the $\sigma$-trace of the first equation above with the second one, one shows that \eqref{sugraEom2} is equivalent to 
\begin{equation}\label{sugraEom3}
i\,\widebar\sigma^\mu\Psi_{\mu\nu A}-\del_\nu\bar\lambda_A=0\;,\quad
i\,\sigma^\mu\del_\mu\bar\lambda_A=0\;,
\end{equation}
and $\sigma^{\mu\nu}\Psi_{\mu\nu A}=0$ is implied by taking the $\sigma$-trace of the spin 3/2 equation.
The dilatino now obeys the Dirac equation, but it is mixed with the string-frame gravitino $\psi_{\mu A}$. To diagonalize them, we perform a linearized super-Weyl transformation and define the Rarita-Schwinger gravitino via
\begin{equation}
    \psi_{\mu A}^{\rm RS} \defeq \psi_{\mu A}-\frac{i}{2}\,\sigma_\mu\,\bar\lambda_A\;,    
\end{equation}
which result in the decoupled equations
\begin{equation}\label{sugraEom4}
    i\,\widebar\sigma^\mu\big(\del_\mu\psi^{\rm RS}_{\nu A}-\del_\nu\psi^{\rm RS}_{\mu A}\big) = 0 \;,\quad
    i\,\sigma^\mu\del_\mu\bar\lambda_A = 0 \;.
\end{equation}
The above equation for the gravitino (together with its complex conjugate), is not the standard Rarita-Schwinger one for a massless spin 3/2 field. Rather, it is the spin 3/2 case of the Fang-Fronsdal equation \cite{Fang:1978wz} for massless fermions of arbitrary spin. We will now show, following \cite{Bouatta:2004kk}, that it is equivalent to the massless Rarita-Schwinger equation. 

In terms of the field strength $\Psi_{\mu\nu }=\del_\mu\psi_{\nu}-\del_\nu\psi_{\mu}$ for a generic spin 3/2 field $\psi_\mu$, the Fang-Fronsdal equation reads $i\,\widebar\sigma^\mu\Psi_{\mu\nu}=0$. The vanishing of its $\sigma$-trace further implies $\sigma^{\mu\nu}\Psi_{\mu\nu}=0$.
We now make use of the identity
\begin{equation}\label{sigmabarsigmaLorentz}
\widebar\sigma^\rho\sigma^{\mu\nu}=\tfrac12\,\epsilon^{\mu\nu\rho\lambda}\,\widebar\sigma_\lambda-i\,\widebar\sigma^{[\mu}\eta^{\nu]\rho}\;,  
\end{equation}
to rewrite the Fang-Fronsdal equation as
\begin{equation}
\begin{split}
0=i\,\widebar\sigma^\mu\Psi_{\mu\nu}&=i\,\widebar\sigma^{[\mu}\delta^{\rho]}_\nu\Psi_{\mu\rho}\\
&=\big(\tfrac12\,\epsilon^{\mu\rho}{}_{\nu\lambda}\,\widebar\sigma^\lambda-\widebar\sigma_\nu\sigma^{\mu\rho}\big)\,\Psi_{\mu\rho}\\
&=\tfrac12\,\epsilon^{\mu\rho}{}_{\nu\lambda}\,\widebar\sigma^\lambda\Psi_{\mu\rho}\;.
\end{split}    
\end{equation}
Written in terms of $\psi_\mu$, the last line is exactly the chiral Rarita-Schwinger equation.

We have thus shown that the field equations \eqref{sugraEom4} are equivalent to
\begin{equation}\label{sugraEom5}
\epsilon^{\mu\nu\rho\sigma}\widebar\sigma_\nu\del_\rho\psi^{\rm RS}_{\sigma A}=0\;,\quad
i\,\sigma^\mu\del_\mu\bar\lambda_A=0\;,
\end{equation}
together with identical equations for $\psi^{\rm RS}_{\mu\tilde A}$ and $\bar\lambda_{\tilde A}$. This concludes the analysis of the NS-R sector, yielding eight gravitini that obey the usual Rarita-Schwinger equation, together with 56 Weyl fermions obeying the chiral Dirac equation.

\subsubsection*{R-R Sector}

Upon solving the section constraint, the Ramond-Ramond bispinors obey the Bargmann-Wigner equations, which are the massless Dirac equation in all spinor indices:
\begin{subequations}\begin{align}
    i\,\widebar\sigma^{\mu\dot\alpha\alpha}\del_\mu \big(F_A{}^{\tilde B}\big)_\alpha{}^{\dot\beta}=i\,\sigma^\mu\!_{\beta\dot\beta}\,\del_\mu \big(F_A{}^{\tilde B}\big)_\alpha{}^{\dot\beta}&=0\;,\\[1mm]
    i\,\widebar\sigma^{\mu\dot\alpha\alpha}\del_\mu \big(F_{A\tilde B}\big)_{\alpha\beta}=i\,\widebar\sigma^{\mu\dot\alpha\alpha}\del_\mu \big(F_{A\tilde B}\big)_{\beta\alpha}&=0\;.    
\end{align}\end{subequations}
Using the decomposition \eqref{RRFierzed}, i.e.
\begin{equation}
    F_A{}^{\tilde B} = \sigma^\mu P_{\mu A}{}^{\tilde B}\;, \quad F_{A\tilde B} = \epsilon\,M_{A\tilde B}+\tfrac12\,\sigma^{\mu\nu}F^+\!\!\!\!\!{}_{\mu\nu A\tilde B} \;, 
\end{equation}
and the identities \eqref{sigmasigmabar}, \eqref{sigmabarsigmaLorentz}, the Bargmann-Wigner equations become 
integrability conditions and Maxwell equations for zero-, one- and two-form field-strengths \begin{subequations}\label{RRMaxwells}\begin{align}
    \del_\mu M_{A\tilde B}&=0\;,\\[1mm]
    \del_{[\mu}P_{\nu]A}{}^{\tilde B}&=0\;,\quad\del^\mu P_{\mu A}{}^{\tilde B}\,=0\;,\\[1mm]
    \del_{[\mu}F^+\!\!\!\!\!{}_{\nu\rho] A\tilde B}&=0\;,\quad \del^\mu F^+\!\!\!\!\!{}_{\mu\nu A\tilde B}=0\;,
\end{align}\end{subequations}
together with identical equations for the complex conjugates.

The vector field strength describes 16 complex scalar fields via
\begin{equation}
    P_{\mu A}{}^{\tilde B}=\del_\mu\Phi_{A}{}^{\tilde B}\;,\quad \Box\Phi_{A}{}^{\tilde B}=0\;.    
\end{equation}
Similarly, the integrability condition for the two-form can be solved by introducing gauge potentials as
\begin{equation}
    F^+\!\!\!\!\!{}_{\mu\nu A\tilde B}=\del_\mu A^+\!\!\!{}_{\nu A\tilde B}-\del_\nu A^+\!\!\!{}_{\mu A\tilde B}\;. 
\end{equation}
However, since the two-form is self-dual, Maxwell's equations are implied by the stronger self-duality constraint
\begin{equation}
    \del_\mu A^+\!\!\!{}_{\nu A\tilde B}-\del_\nu A^+\!\!\!{}_{\mu A\tilde B}=-i\,\epsilon_{\mu\nu}{}^{\rho\sigma}\del_\rho A^+\!\!\!{}_{\sigma A\tilde B} \;,   
\end{equation}
which is to be viewed as the independent dynamical equation for the gauge fields $A^+\!\!\!{}_{\mu A\tilde B}$. In the same manner, the complex conjugate two-form $\bar F_{\mu\nu}^{-\,A\tilde B}$ is related to the anti-self-dual gauge fields $\bar A_{\mu}^{-\,A\tilde B}\equiv\big(A^+\!\!\!{}_{\mu A\tilde B}\big)^*$. This means that $A^+\!\!\!{}_{\mu A\tilde B}$ propagates 16 helicity $+1$ eigenstates, while $\bar A_{\mu}^{-\,A\tilde B}$ propagates the 16 helicity $-1$ CPT conjugate states. As we have anticipated, together they count for the on-shell degrees of freedom of 16 real vector fields. However, their self-dual and anti-self-dual parts transform in inequivalent representations of $SU(4)\times SU(4)$. We will discuss this further in the next subsection. 

Finally, the scalar matrix $M_{A\tilde B}$ is constant on-shell and represents 16 non-propagating degrees of freedom, which we can in principle discard. Nevertheless, one might observe that in four dimensions we can either dualize the scalar field-strength $M_{A\Btilde}$ into the four-form 
\begin{equation}
    M_{A\Btilde} = \frac{1}{4!}\epsilon^{\mu\nu\rho\sigma}F_{\mu\nu\rho\sigma A\Btilde} \,,
\end{equation}
which is the field-strength of a topological $3$-form potential $C_{\mu\nu\rho A\tilde B}$, or consider the non-propagating zero-form $M_{A\Btilde}$ as a constant mass term. Either way, keeping this field strength is equivalent to including massive deformations in the theory, analogous to those obtained  from the compactification of $D=11$ supergravity by keeping the 3-form in four dimensions \cite{Aurilia:1980xj}.

\subsection{Enhancement of $R$-symmetry and Supersymmetry}
\label{Subsec:Sym_Enhancement}

One last test to verify that what we have attained actually is a DFT version of $\calN=8$ supergravity, consists in proving that the double copy theory shows the $R$-symmetry enhancement 
\begin{equation}
    SU(4) \times SU(4) \to SU(8)\;,
\end{equation}
upon taking the supergravity solution of the section constraint. As for the counting, in section \ref{Subsec:DC_Global_SUSY} we have already shown that the double copy is invariant under eight global supersymmetries, in terms of the doubled parameter
\begin{equation}\label{doubledparameter}
    \epsilon_I \defeq (\epsilon_A, \epsilon_{\tilde A})\;, \quad I=1,\dots,8 \;,
\end{equation}
together with the supersymmetry $n$-linear maps $\bbSigma_n(\epsilon)$.

As suggested by \eqref{doubledparameter}, we define the index $I\defeq (A,\tilde A)$, $I=1,\ldots,8$ in the fundamental of $SU(8)$. We will then show that the field content can be organized in terms of $SU(8)$ representations. This is a non-trivial statement already at the free level, since the double copy naturally exhibits an $SU(4)\times SU(4)$ $R$-symmetry, with fundamental indices $A=1,\ldots,4$ and $\tilde A=1,\ldots,4$. Besides the graviton, which is obviously an $SU(8)$ singlet, we start our analysis from the gravitini.

\subsubsection*{Gravitini}

The eight gravitini are the defining feature of $\cN=8$ supergravity. The left-handed $\psi^{\rm RS}_{\mu A}$ and $\psi^{\rm RS}_{\mu \Atilde}$ arrange naturally into the fundamental representation ${\bf 8}$ of $SU(8)$:
\begin{equation}
    \psi_{\mu I} \defeq \big(\psi^{\rm RS}_{\mu A}, \psi^{\rm RS}_{\mu \Atilde}\big)\;,
\end{equation}
while their complex conjugates  are right-handed and belong to the anti-fundamental $\bar{\bf 8}$
\begin{equation}
\widebar\psi_\mu^I\defeq \big(\widebar\psi_\mu^{{\rm RS}A},\widebar\psi_\mu^{{\rm RS}\tilde A}\big)\equiv(\psi_{\mu I})^\dagger\;.    
\end{equation}

\subsubsection*{Vector Fields}

For the vector fields the situation is more subtle. We find it more transparent to work with their abelian field strengths
\begin{equation}
    F_{\mu\nu AB}\;,\quad F_{\mu\nu \tilde A\tilde B}\;,\quad F^+\!\!\!\!\!{}_{\mu\nu A\tilde B}\;,\quad \bar F_{\mu\nu}^{-\,A\tilde B}\;,    
\end{equation}
where we recall that the first two come from the NS-NS sector, while the self-dual and anti-self-dual ones come from the R-R sector.
According to \eqref{Eq:Reality_Cond_Scalars} and \eqref{Eq:Reality_Cond_Spinors}, they are subject to the reality conditions
\begin{subequations}\begin{align}
    \big(F_{\mu\nu AB}\big)^*&=\bar F_{\mu\nu}{}^{AB}=\tfrac12\,\epsilon^{ABCD}F_{\mu\nu CD}\;,\\[1mm]
    \big(F_{\mu\nu \tilde A\tilde B}\big)^*&=\bar F_{\mu\nu}{}^{\tilde A\tilde B}=\tfrac12\,\epsilon^{\tilde A\tilde B\tilde C\tilde D}F_{\mu\nu \tilde C \tilde D}\;,\\[1mm]
    \big(F^+\!\!\!\!\!{}_{\mu\nu A\tilde B}\big)^*&=\bar F_{\mu\nu}^{-\,A\tilde B}\;.
\end{align}\end{subequations}
In order to collect them in a consistent $SU(8)$ representation, we first split the NS-NS field strengths into their self-dual and anti-self-dual parts 
(c.f.~(\ref{Fpmeq})):
\begin{equation}
    F_{\mu\nu AB}=F^{+}_{\mu\nu AB}+F^{-}_{\mu\nu AB}\;,\quad F_{\mu\nu \tilde A \tilde B}=F^{+}_{\mu\nu \tilde A\tilde B}+F^{-}_{\mu\nu \tilde A\tilde B} \;,
\end{equation}
obeying the reality conditions
\begin{subequations}\begin{align}
    \big(F^{+}_{\mu\nu AB}\big)^*&=\bar F^{-AB}_{\mu\nu}=\tfrac12\,\epsilon^{ABCD}F^-_{\mu\nu CD}\;,\\[1mm]
    \big(F^{+}_{\mu\nu \tilde A\tilde B}\big)^*&=\bar F^{-\tilde A\tilde B}_{\mu\nu}=\tfrac12\,\epsilon^{\tilde A\tilde B\tilde C\tilde D}F^-_{\mu\nu \tilde C\tilde D}\;.
\end{align}\end{subequations}
This shows that the full NS-NS field strengths can be represented by their self-dual parts $F^+_{\mu\nu AB}$ and $F^+_{\mu\nu \tilde A\tilde B}$, together with their complex conjugates.
At this point, we can collect all the self-dual field strengths into the {\bf 28} antisymmetric tensor representation of $SU(8)$
\begin{equation}
F^+_{\mu\nu\, IJ}\defeq \big(F^+_{\mu\nu\, AB}\,,\,F^+_{\mu\nu\, \tilde A \tilde B}\,,\,F^+_{\mu\nu\, A\tilde B}\big)\;,    
\end{equation}
while their anti-self-dual complex conjugates $\bar F_{\mu\nu}^{-IJ}$ belong to the $\widebar{\bf 28}$.

\subsubsection*{Spin 1/2 fermions}

The left-handed Weyl fermions of the theory are given by
\begin{equation}
\chi_{\Atilde BC}\;,\quad \chi_{A\Btilde\Ctilde}\;,\quad \lambda^{A}\;,\quad \lambda^{\Atilde}\;.
\end{equation}
The dilatini can be dualized into
\begin{equation}
\chi_{ABC}\defeq \epsilon_{ABCD}\lambda^D\;,\quad\chi_{\tilde A\tilde B\tilde C}\defeq \epsilon_{\tilde A\tilde B\tilde C\tilde D}\lambda^{\tilde D}\;,   
\end{equation}
using the $SU(4)$ epsilon tensors.
This allows us to group all spin 1/2 fermions into the rank-three antisymmetric tensor representation ${\bf 56}$ of $SU(8)$ by defining
\begin{equation}
    \chi_{IJK} \defeq \big(\chi_{\Atilde BC}\,,\, \chi_{A\Btilde\Ctilde}\,,\, \chi_{ABC}\,,\, \chi_{\Atilde\Btilde\Ctilde}\big)\;,
\end{equation}
which is the standard representation of the Weyl fermions in $\cN=8$ supergravity. The right-handed complex conjugate $\bar\chi^{IJK}$ transform in the $\widebar{\bf 56}$.

\subsubsection*{Scalar Fields}

The scalar fields of the theory are the 36 $\Phi_{AB\tilde C\tilde D}$, together with the dilaton $\phi$ and axion $a$, from the NS-NS sector, while the R-R sector provides 16 more complex scalars $\Phi_A{}^{\tilde A}$. They obey the reality conditions
\begin{subequations}\begin{align}
    \big(\Phi_{AB\tilde C\tilde D}\big)^*&=\widebar{\Phi}^{AB\tilde C\tilde D}=\frac14\,\epsilon^{ABCD}\epsilon^{\tilde A\tilde B\tilde C\tilde D}\Phi_{CD\tilde A\tilde B}\;,\\[1mm]
    \big(\Phi_A{}^{\tilde B}\big)^*&=\widebar{\Phi}^A{}_{\tilde B}\;,
\end{align}\end{subequations}
while the axion and dilaton are real. This suggests defining
\begin{subequations}\begin{align}
\Phi_{ABCD} &\defeq \epsilon_{ABCD}(\phi + ia)\;, \\[1mm]
    \Phi_{\Atilde\Btilde\Ctilde\Dtilde} &\defeq \epsilon_{\Atilde\Btilde\Ctilde\Dtilde}(\phi - ia)\;, \\[1mm]
    \Phi_{A\Btilde\Ctilde\Dtilde} &\defeq \epsilon_{\Btilde\Ctilde\Dtilde\tilde E}\,\Phi_A{}^{\tilde E}  \;,\\[1mm]
    \Phi_{ABC\Dtilde} &\defeq \epsilon_{ABCD}\widebar{\Phi}^D{}_\Dtilde\;. 
\end{align}\end{subequations}
In this way, the scalars fill the rank-four antisymmetric tensor representation ${\bf 70}$ of $SU(8)$
\begin{equation}
    \Phi_{IJKL} \defeq \big(\Phi_{ABCD}, \Phi_{\Atilde\Btilde\Ctilde\Dtilde}, \Phi_{ABC\Dtilde}, \Phi_{A\Btilde\Ctilde\Dtilde}, \Phi_{AB\Ctilde\Dtilde}\big)\;,
\end{equation}
subject to the reality condition
\begin{equation}
\big(\Phi_{IJKL}\big)^*=\widebar{\Phi}^{IJKL}=\frac{1}{4!}\,\epsilon^{IJKLMNPQ}\Phi_{MNPQ}\;,    
\end{equation}
which describes 70 real scalar degrees of freedom.

%%%%%%%%%%%%%%%%%%%%%%%%%%%%%%%%%%
%%%%%%%%%%%%%%%%%%%%%%%%%%%%%%%%
%% BEGIN SECTION 5 (CONCL.) %%%%%%
%%%%%%%%%%%%%%%%%%%%%%%%%%%%%%%%
%%%%%%%%%%%%%%%%%%%%%%%%%%%%%%%%%%
\section{Conclusions and Outlook}
\label{Sec:Conclusions}

In this paper we have applied the program of finding a first-principle derivation of the double copy via homotopy algebras to the maximally supersymmetric theories in $D=4$, with the goal to realize ${\cal N}=8$ supergravity, in an off-shell, gauge invariant and local formulation, as the double copy of ${\cal N}=4$ SYM. 
We have shown that, at least to cubic order in fields, the $L_{\infty}$ algebra of ${\cal N}=8$ supergravity can be obtained by double copying the kinematic algebra ${\cal K}$ of ${\cal N}=4$ SYM, using a redundant formulation for the fermionic fields. 
Most importantly, in this we have shown structurally how the ${\cal N}=4$ global supersymmetry of SYM gets double copied to the ${\cal N}=8$ supersymmetry of maximal supergravity.  

These results are just a small part  of an ongoing research program to unveil a deep connection between gauge theory and gravity that promises to be particularly powerful in the realm of maximally supersymmetric theories. In particular, this program should be continued along the following directions: 
\begin{itemize}

    \item Arguably the most urgent outstanding problem is to extend the construction to all orders in fields. The double copy of pure Yang-Mills theory has been pushed to quartic order in fields, and it has been shown that the $L_{\infty}$ brackets of the gravity theory can be constructed purely in terms of the BV$_{\infty}^{\Box}$ structure maps of the kinematic algebra of Yang-Mills theory \cite{Bonezzi:2022bse}. 
    It remains to generalize the double copy of the ${\cal N}=4$ global supersymmetries to these maps. More generally, one needs to find an all-order prescription, ideally through some sort of derived construction from a simpler and possibly strict algebra. 
    Here the recent investigation in terms of vertex operators and the strict operator algebra on the Hilbert space of a worldline theory seems promising \cite{Bonezzi:2024fhd}. 
    Such an improved understanding would also be necessary in order to double copy exact but perturbative solutions of classical (super-)Yang-Mills theory to obtain solutions of (super-)gravity.  
    
    \item It would be interesting to explore supersymmetric theories in other dimensions, with different numbers of supercharges and possibly different combinations of supersymmetric gauge theories, as in the general web of theories explored in the literature \cite{Bern:2019prr, Anastasiou:2015vba}. 
    A particularly interesting example is of course $D=10$, where the double copy of ${\cal N}=1$ super-Yang-Mills theory should lead to ${\cal N}=2$ type IIA or IIB supergravity in a double field theory formulation as in \cite{Hohm:2011zr,Hohm:2011dv}. 
    Moreover, the fate of other global symmetries on the gauge theory side should be investigated further. 
    For instance, ${\cal N}=4$ SYM features, in addition to its super-Poincar\'e symmetries, super-conformal transformations that do not appear to have a counterpart in the double copied ${\cal N}=8$ supergravity, but it might be beneficial to explore whether they play a hidden role. Finally, it would be interesting to understand the global U-duality E$_{7(7)}$ symmetry of ${\cal N}=8$ supergravity in terms of the double copy construction. 
    
    \item One shortcoming of the present construction, owing to the redundant formulation for the fermionic fields, is that the double copy is manifestly non-Lagrangian, featuring a different number of fields and field equations. It will be important to generalize this construction in order to allow for an action principle. One manifestation of the non-Lagrangian character is the Ramond-Ramond sector in which not the gauge fields but rather their field strengths appear as the elementary fields. In fact, the problem of writing an action including the Ramond-Ramond sector is closely related to the corresponding problem in type II string field theory, where a resolution was proposed by Sen a few years ago \cite{Sen:2015nph}, and it would be interesting to investigate whether this formulation could be useful in the double copy context. 
    
    \item One of the most exciting prospects of the double copy relation between ${\cal N}=4$ SYM and ${\cal N}=8$ supergravity has always been that it might give one an instrument to settle the problem whether ${\cal N}=8$ supergravity could be UV finite. 
    While the UV-finiteness of ${\cal N}=4$ SYM  was proved in the early 1980s  \cite{Brink:2015ust}, the UV-finiteness of ${\cal N}=8$ supergravity has remained open ever since its discovery.
    With no hope in sight of settling the issue, this is a bit of an embarrassment, and one might wonder whether a fundamentally different approach towards quantum field theory and gravity is needed. 
    The homotopy algebra approach to field theory, and the closely related program of Costello and Gwilliam to formulate quantum field theory in terms of factorization algebras \cite{Costello:2016vjw,Costello:2021jvx}, might turn out to be an important ingredient towards the goal of uncovering the UV properties of ${\cal N}=8$ supergravity. 

\end{itemize}

%%%%%%%%%%%%%%%%%%%%%%%%%%%%%%%%%%
%%%%%%%%%%%%%%%%%%%%%%%%%%%%%%%%
%% BEGIN ACKNOWLEDGEMENTS %%%%%%%%
%%%%%%%%%%%%%%%%%%%%%%%%%%%%%%%%
%%%%%%%%%%%%%%%%%%%%%%%%%%%%%%%%%%
\section*{Acknowledgments}
We would like to thank Maor Ben-Shahar, Christoph Chiaffrino, Felipe D\'iaz-Jaramillo, Georgios Itsios and Maria Kallimani for discussions and collaborations on closely related topics.

The work of R.B.~is funded by the Deutsche Forschungsgemeinschaft (DFG, German Research
Foundation) -- Projektnummer 524744955, ``Worldline approach to the double copy", and G.C.~is funded by the DFG -- Projektnummer 417533893/GRK2575 “Rethinking Quantum Field Theory”.

\appendix

%%%%%%%%%%%%%%%%%%%%%%%%%%%%%%%%%%
%%%%%%%%%%%%%%%%%%%%%%%%%%%%%%%%
%% BEGIN APPENDIX A %%%%%%%%%%%%%%
%%%%%%%%%%%%%%%%%%%%%%%%%%%%%%%%
%%%%%%%%%%%%%%%%%%%%%%%%%%%%%%%%%%
\section{Conventions and Properties}
\label{Sec:Conventions}

We pick a mostly-minus signature, such that the four-dimensional Minkowski metric reads
\begin{equation}
    \eta_{\mu\nu} ={\rm diag} (1,-1,-1,-1) \,,
\end{equation}
with $\mu,\nu=0,\dots,3$. The totally antisymmetric Levi-Civita tensor $\epsilon_{\mu\nu\rho\sigma}$ is defined as
\begin{equation}
    \epsilon^{0123} = -\epsilon_{0123} = 1\,, \quad \epsilon^{\mu\nu\rho\sigma}\epsilon_{\mu\nu\rho\sigma} = -4!\;,
\end{equation}
while the Hodge duality operator $\star$ acts on a $p$-form $\omega$ as
\begin{equation}
    (\star\omega)_{\mu_1\dots\mu_{4-p}} \defeq \frac{1}{p!}\epsilon_{\mu_1\dots\mu_{4-p}\nu_1\dots\nu_{p}}\omega^{\nu_1\dots\nu_{p}} \,.
\end{equation}
Finally, the self-dual and anti-self-dual parts of a two-form $F_{\mu\nu}$ are defined by
\begin{equation}\label{Fpmeq}
    F^{\pm}_{\mu\nu} = \frac{1}{2}\left(F_{\mu\nu} \mp i \star\!F_{\mu\nu}\right) = \frac{1}{2}\left(F_{\mu\nu} \mp \frac{i}{2}\epsilon_{\mu\nu}{}^{\rho\sigma}F_{\rho\sigma}\right)\;,
\end{equation}
such that any two-form can be decomposed as $F_{\mu\nu}=F^+_{\mu\nu}+F^-_{\mu\nu}$, where
\begin{equation}
    \star F^\pm_{\mu\nu} = \pm\, iF^\pm_{\mu\nu}\;.
\end{equation}
%%%%%%%%%%%%%%%%%%%%%%%%%%%%%%%%%
%%%%%%%%%%%%%%%%%%%%%%%%%%%%%%%%
%%%%%%%%%%%%%%%%%%%%%%%%%%%%%%%%%%
\subsubsection*{Spinor conventions}

As usual, we denote left-handed spinor indices by $\alpha,\beta,\cdots=1,2$ and right-handed spinor indices by $\dot\alpha,\dot\beta,\ldots=\dot1,\dot2$.
The $SL(2,\mathbb{C})$ invariant symbols are defined as
\begin{equation}
    \epsilon^{\alpha\beta} = \epsilon^{\dot\alpha\dot\beta} \defeq \begin{pmatrix}
        0 & 1 \\ -1 & 0
    \end{pmatrix}, \quad
    \epsilon_{\alpha\beta} = \epsilon_{\dot\alpha\dot\beta} \defeq \begin{pmatrix}
        0 & -1 \\ 1 & 0
    \end{pmatrix}
\end{equation}
so that we can raise and lower the indices following the rule
\begin{equation}
    \lambda^\alpha \defeq \epsilon^{\alpha\beta}\lambda_\beta \,, \quad \lambda_\alpha \defeq \epsilon_{\alpha\beta}\lambda^\beta \,.
\end{equation}
The indices of left-handed and right-handed spinors are contracted respectively as
\begin{equation}
    \lambda\,\xi \defeq \lambda^\alpha\,\xi_\alpha, \quad \widebar\lambda\,\widebar\xi \defeq \widebar\lambda_{\dot\alpha}\,\widebar\xi^{\dot\alpha}\,.
\end{equation}
For the $SL(2,\mathbb{C})$ sigma matrices we take 
\begin{equation}
(\sigma^\mu)_{\alpha\dot\alpha} = (\mathbbm{1};\Vec{\sigma})_{\alpha\dot\alpha}\;,\quad   (\widebar\sigma^\mu)^{\dot\alpha\alpha} = (\mathbbm{1};-\Vec{\sigma})^{\dot\alpha\alpha}\;, 
\end{equation}
where $\Vec{\sigma}\equiv(\sigma^1,\sigma^2,\sigma^3)$ are the standard Pauli matrices. The Lorentz generators for the left-handed and right-handed spinor representations are given by
\begin{equation}
    (\sigma^{\mu\nu})_{\alpha}{}^{\beta} \defeq \frac i4\left( \sigma^\mu\widebar\sigma^\nu - \sigma^\nu\widebar\sigma^\mu \right)_{\alpha}{}^{\beta} \,, \quad (\widebar\sigma^{\mu\nu})^{\dot\alpha}{}_{\dot\beta} \defeq \frac i4 \left(\widebar\sigma^\mu\sigma^\nu - \widebar\sigma^\nu\sigma^\mu \right)^{\dot\alpha}{}_{\dot\beta} \,,
\end{equation}
and obey the (anti)-self-duality conditions
\begin{equation}\begin{split}
    \sigma^{\mu\nu} = -\frac i2\epsilon^{\mu\nu\rho\sigma}\sigma_{\rho\sigma} \,, \quad \widebar\sigma^{\mu\nu} = \frac i2\epsilon^{\mu\nu\rho\sigma}\widebar\sigma_{\rho\sigma} \,.
\end{split}\end{equation}
We list here some useful identities for products of sigma matrices:
\begin{subequations}\begin{align}
    (\sigma^\mu)_{\alpha\dot\alpha}(\widebar\sigma_\mu)^{\dot\beta\beta} &= 2\delta_{\alpha}^{\beta}\delta_{\dot\alpha}^{\dot\beta}\;, \\[1mm]
    (\sigma^{\mu\nu})_\alpha{}^\beta(\sigma_{\mu\nu})_\gamma{}^\delta &= \epsilon_{\alpha\gamma}\epsilon^{\beta\delta} + \delta_\alpha^\delta\delta_\gamma^\beta\;, \\[1mm]
    \sigma^\mu\widebar\sigma^\nu &= \eta^{\mu\nu} - 2i\sigma^{\mu\nu}\;, \\[1mm]
    \widebar\sigma^\mu\sigma^\nu &= \eta^{\mu\nu} - 2i\widebar\sigma^{\mu\nu}\;, \\[1mm]
    \sigma^{\mu\nu}\sigma^\rho &= - \frac12\epsilon^{\mu\nu\rho\gamma}\sigma_\gamma + i\sigma^{[\mu}\eta^{\nu]\rho}\;, \\[1mm]
    \widebar\sigma^{\mu\nu}\widebar\sigma^\rho &= + \frac12\epsilon^{\mu\nu\rho\gamma}\widebar\sigma_\gamma + i\widebar\sigma^{[\mu}\eta^{\nu]\rho}\;, \\[1mm]
    \sigma^\rho\widebar\sigma^{\mu\nu} &= - \frac12\epsilon^{\mu\nu\rho\gamma}\sigma_\gamma - i\sigma^{[\mu}\eta^{\nu]\rho}\;, \\[1mm]
    \widebar\sigma^\rho\sigma^{\mu\nu} &= + \frac12\epsilon^{\mu\nu\rho\gamma}\widebar\sigma_\gamma - i\widebar\sigma^{[\mu}\eta^{\nu]\rho} \;,\\[1mm]
    \sigma^\mu\widebar\sigma^\nu\sigma^\rho - \sigma^\rho\widebar\sigma^\nu\sigma^\mu &= +2i \epsilon^{\mu\nu\rho\sigma}\sigma_\sigma\;, \\[1mm]
    \widebar\sigma^\mu\sigma^\nu\widebar\sigma^\rho - \widebar\sigma^\rho\sigma^\nu\widebar\sigma^\mu &= -2i \epsilon^{\mu\nu\rho\sigma}\widebar\sigma_\sigma \;.
\end{align}\end{subequations}

%%%%%%%%%%%%%%%%%%%%%%%%%%%%%%%%%%
%%%%%%%%%%%%%%%%%%%%%%%%%%%%%%%%
%% BEGIN APPENDIX B %%%%%%%%%%%%%%
%%%%%%%%%%%%%%%%%%%%%%%%%%%%%%%%
%%%%%%%%%%%%%%%%%%%%%%%%%%%%%%%%%%
\section{Fermionic complex from the spinning particle}
\label{Sec:Fermionic_Complex}

In this appendix we show that the fermionic complex employed in section \ref{Sec:Hom_Alg_Of_SYM} is generated by the BRST quantization of the relativistic spinning particle \cite{Berezin:1976eg,Brink:1976sz,Barducci:1976qu}.

The first-quantized theory describing spin one-half fermions in target space is the relativistic particle with $N=1$ real supersymmetry on the worldline. The action in phase space reads
\begin{equation}\label{susyparticle}
    S=\int d\tau\,\Big[p_\mu\dot x^\mu+\tfrac{i}{2}\,\psi^\mu\dot\psi_\mu-\tfrac12\,e\,p^2-i\,\chi\,\psi^\mu p_\mu\Big]\;,    
\end{equation}
where $x^\mu(\tau)$ and $p^\mu(\tau)$ are the four-position and four-momentum of the particle, while $\psi^\mu(\tau)$ are real Grassmann variables transforming as vectors under the target space Lorentz group $SO(1,3)$. The Lagrange multipliers $e(\tau)$ and $\chi(\tau)$ play the role of one-dimensional einbein and gravitino, respectively. Besides manifest Poincar\'e symmetry in target space, the action \eqref{susyparticle} is invariant under time reparametrizations with parameter $\xi(\tau)$ and local worldline supersymmetry, with Grassmann odd parameter $\epsilon(\tau)$:  
\begin{subequations}\begin{align}
    \delta x^\mu&=\xi p^\mu+i\,\epsilon\psi^\mu\;, \hspace{1cm}
    \delta\psi^\mu=-\epsilon p^\mu\;,\\[1mm]
    \delta e&=\dot\xi+2i\,\epsilon\chi\;, \hspace{1.65cm} \delta\chi=\dot\epsilon\;,
\end{align}\end{subequations}
while $p_\mu$ is gauge invariant.
The above transformations are canonically generated by the Hamiltonian $H=\frac12\,p^2$ and supercharge $G=\psi^\mu p_\mu$.

Upon canonical quantization, the phase space variables obey the following commutation relations:
\begin{equation}
[x^\mu,p_\nu]=i\,\delta^\mu_\nu\;,\quad\{\psi^\mu,\psi^\nu\}=\eta^{\mu\nu}\;.    
\end{equation}
While the $(x,p)$ algebra is realized as usual on functions of $x^\mu$, with $p_\mu\equiv-i\del_\mu$, the quantum Grassmann variables obey a Clifford algebra. This can be realized by taking spacetime spinor fields as states in the Hilbert space, with the identification $\psi^\mu=\frac{1}{\sqrt2}\,\Gamma^\mu$.
To comply with the main text, we can fix the spacetime dimension to be four and take the Dirac matrices $\Gamma^\mu$ in the chiral basis, where they are given by
\begin{equation}
\Gamma^\mu=\bpm0&\sigma^\mu\\\bar\sigma^\mu&0\epm\;.    
\end{equation}
The classical equations of motion for the einbein $e$ and gravitino $\chi$ set the Hamiltonian and supercharge to zero. In the quantum theory, these turn into an algebra of first-class constraints
\begin{equation}\label{susyconstraintparticle}
\{G,G\}=2H\;,\quad [G,H]=0 \;,  
\end{equation}
which is the $N=1$ supersymmetry algebra in one dimension. The role of the above constraints in the quantum theory is to impose the mass-shell condition and the massless Dirac equation on physical states.

\subsubsection*{BRST quantization}

In the quantum theory, the first-class constraints \eqref{susyconstraintparticle} can be treated in the BRST framework. To this end, one introduces a ghost-antighost canonical pair for each constraint:
\begin{subequations}\begin{align}
    H\;\longrightarrow\;(b,c)\;,\,\quad \{b,c\}&=1\;,\\[1mm] 
    G\;\longrightarrow\;(\beta,\gamma)\;,\quad [\beta,\gamma]&=1\;,
\end{align}\end{subequations}
with $(b,c)$ Grassmann odd, so that $b^2=0$ and $c^2=0$, and $(\beta,\gamma)$ Grassmann even. We assign ghost number $+1$ to $(c,\gamma)$ and $-1$ to the antighosts $(b,\beta)$. We choose the ghost Hilbert space to be generated by monomials of arbitrary degree in $\gamma$ and $c$, so that a generic state in the BRST Hilbert space $\cH$ is given by a spacetime spinor $\Psi(x,\gamma,c)$, where the dependence on the bosonic ghost $\gamma$ is intended to be polynomial. The momentum operator and the antighosts act on this Hilbert space as the derivative operators
\begin{equation}
    p_\mu\equiv - i\del_\mu \;, \quad
    b\equiv\frac{\del}{\del c}\;,\quad\beta\equiv\frac{\del}{\del\gamma}\;.
\end{equation}
A nilpotent BRST operator of ghost number $+1$ is given by
\begin{equation}
    Q = c\,\Box + i\gamma\delslash + \gamma^2\frac{\del}{\del c}\;,  \quad Q^2=0\;,
\end{equation}
where ${\Box=\del^\mu\del_\mu}$ and ${\delslash=\Gamma^\mu\del_\mu}$ are the massless Klein-Gordon and Dirac operators, respectively. We will now show that the BRST Hilbert space coincides with the modified fermionic complex \eqref{CC:SYM_Linf_ferm}, with differential $B_1\equiv Q$.

Since a generic state in the Hilbert space $\cH$ is given by a spinor $\Psi(x,\gamma,c)$ with non-negative ghost number, elements of homogeneous ghost number decompose the Hilbert space into the semi-infinite complex
\begin{equation}
    \cH = \bigoplus_{n=0}^\infty\cH_n\;,\quad Q:\cH_n\longrightarrow\cH_{n+1}\;.
\end{equation}
A state in ghost number zero consists of a single fermion $\psi(x)$, to be identified as the field. At every nonzero ghost number the space contains a doublet of spinors, namely
\begin{equation}
    \Psi_n(x,\gamma,c) = \gamma^n\,\psi_n(x)+c\,\gamma^{n-1}\chi_n(x)\;.    
\end{equation}
This decomposition of the Hilbert space shows that it is isomorphic to the fermionic complex\footnote{Here we are working with Dirac spinors with no $R$-symmetry. To recover the formulation of the main text one has to first split the fermion into left- and right-handed Weyl spinors.} \eqref{CC:SYM_Linf_ferm}, upon identifying the $L_\infty$ degree with the worldline ghost number: $X^{\rm F}_n\simeq\cH_n$. To see that the differential $B_1$ is to be identified with the BRST operator $Q$, we take a field $\psi(x)\in\cH_0$ and compute
\begin{equation}\label{Qpsi}
    Q\psi = i\gamma\,\delslash\psi + c\,\Box\psi \; \in\cH_1\;,
\end{equation}
which yields both the Dirac and the redundant Klein-Gordon equations, as in \eqref{CC:SYM_Linf_ferm}. Using the notation of section \ref{Sec:Hom_Alg_Of_SYM}, a state at ghost number $+1$ is written as
\begin{equation}
\Psi_1(x,\gamma,c)=\gamma\,\cE(x)+c\,E(x)\;,    
\end{equation}
and describes the doublet of equations arising from \eqref{Qpsi}. The redundancy of the second-order equation is then encoded in the space of Noether identities $\cH_2$, where
\begin{equation}
Q\Psi_1=\gamma^2\,(i\delslash\cE+E)+c\gamma\,(\Box\cE-i\delslash E)\;\in\cH_2\;,    
\end{equation}
which is the same action as the differential $B_1$ of \eqref{CC:SYM_Linf_ferm}.
The two components above are also not independent, thus generating a Noether-for-Noether identity. This pattern continues ad infinitum, which is expected from the nilpotent operator $Q$ acting on a space with ghost number unbounded from above. Finally, the $b$ operator of the fermionic complex described in section \ref{Sec:Hom_Alg_Of_SYM} is precisely the antighost $b$ of the worldline theory.

%%%%%%%%%%%%%%%%%%%%%%%%%%%%%%%%%%
%%%%%%%%%%%%%%%%%%%%%%%%%%%%%%%%
%% BEGIN BIBLIOGRAPHY %%%%%%%%%%%%
%%%%%%%%%%%%%%%%%%%%%%%%%%%%%%%%
%%%%%%%%%%%%%%%%%%%%%%%%%%%%%%%%%%
\providecommand{\href}[2]{#2}\begingroup\raggedright\endgroup

\end{document}